\documentclass[a4paper,10pt]{article}
\usepackage{graphicx}
\usepackage{comment}
\usepackage{amssymb}
\usepackage{amsmath}
\usepackage{nicefrac}
\usepackage{graphicx}
\usepackage{dcolumn}
\usepackage{bm}
\usepackage{comment}
\usepackage{multirow}
\usepackage{cite}
\usepackage{xcolor}
\usepackage{url}
\usepackage{mathrsfs}
\usepackage{supertabular}
\usepackage[normalem]{ulem}
\usepackage{lscape}
\usepackage{soul}
\usepackage{subcaption} 
\usepackage{bold-extra}
\usepackage{hyperref}
\usepackage{subcaption}
\usepackage{braket}
\usepackage{ulem}
\usepackage{tabularx} 
\usepackage{psfrag} 
\usepackage{sistyle} 

\begin{document}
            
\newcommand{\bin}[2]{\left(\begin{array}{c}\!#1\!\\\!#2\!\end{array}\right)}
\newcommand{\threej}[6]{\left(\begin{array}{ccc}#1 & #2 & #3 \\ #4 & #5 & #6 \end{array}\right)}
\newcommand{\sixj}[6]{\left\{\begin{array}{ccc}#1 & #2 & #3 \\ #4 & #5 & #6 \end{array}\right\}}
\newcommand{\regge}[9]{\left[\begin{array}{ccc}#1 & #2 & #3 \\ #4 & #5 & #6 \\ #7 & #8 & #9 \end{array}\right]}
\newcommand{\La}[6]{\left[\begin{array}{ccc}#1 & #2 & #3 \\ #4 & #5 & #6 \end{array}\right]}
\newcommand{\hj}{\hat{J}}
\newcommand{\hux}{\hat{J}_{1x}}
\newcommand{\hdx}{\hat{J}_{2x}}
\newcommand{\huy}{\hat{J}_{1y}}
\newcommand{\hdy}{\hat{J}_{2y}}
\newcommand{\huz}{\hat{J}_{1z}}
\newcommand{\hdz}{\hat{J}_{2z}}
\newcommand{\hup}{\hat{J}_1^+}
\newcommand{\hum}{\hat{J}_1^-}
\newcommand{\hdp}{\hat{J}_2^+}
\newcommand{\hdm}{\hat{J}_2^-}

\huge

\begin{center}
Modeling stopping power of ions in plasmas using parametric potentials
\end{center}

\vspace{0.5cm}

\large

\begin{center}
Tanguy Barges Delattre$^a$, Sébastien Rassou$^{b,c}$ and Jean-Christophe Pain$^{b,c,}$\footnote{jean-christophe.pain@cea.fr}
\end{center}

\normalsize

\begin{center}
\it $^a$École Normale Supérieure de Lyon, CNRS, Laboratoire de Physique, F-69342 Lyon, France
\it $^b$CEA, DAM, DIF, F-91297 Arpajon, France\\
\it $^b$Universit\'e Paris-Saclay, CEA, Laboratoire Mati\`ere en Conditions Extr\^emes,\\
\it 91680 Bruy\`eres-le-Ch\^atel, France\\
\it $^c$Laboratoire Aim\'e Cotton, Universit\'e Paris-Saclay, Orsay, France\\
\end{center}

\vspace{0.5cm}

\begin{abstract}
We present a study of the ion stopping power due to free and bound electrons in a warm dense plasma. Our main goal is to propose a method of stopping-power calculation expected to be valid for any ionization degree. The free-electron contribution is described by the Maynard-Deutsch-Zimmerman formula and the bound-electron one relies on the Bethe formula with corrections, in particular taking into account density and shell effects. The impact of the bound-state computation by three different parametric potentials is investigated within the Garbet formalism for the mean excitation energy. The first parametric potential is due to Green, Sellin and Zachor, the second one was proposed by Yunta, and the third one was introduced by Klapisch in the framework of atomic-structure computations. The results are compared to the ones of self-consistent average-atom calculations. This approach correctly bridges the limits of neutral and fully-ionized matter. 
\end{abstract}

\section{Introduction}\label{intro}

Due to interactions with matter, ions tend to lose kinetic energy when passing through materials. The loss of energy per unit of length traveled by the ion is called the stopping power of the material. Stopping powers find applications in various fields of research, from thermonuclear fusion to proton therapy. Thermonuclear fusion requires heating a plasma to extreme temperatures (around $10^8$K), densities (a few hundreds of g/cm$^3$) and pressures (a few hundreds of Mbar) in order to overcome the Coulomb barrier. Such conditions are comparable to the ones of stellar interiors. In stars, however, the plasma is confined by gravity. In the facilities dedicated to fusion research, the plasma can be confined by strong magnetic fields \cite{Ongena2016}, as with the ITER (International Thermonuclear Experimental Reactor) project, or by intense-laser irradiation (Inertial Confinement Fusion, ICF) with the NIF (National Ignition Facility) in California and the LMJ (Laser M\'ega Joule) in France. The concept of ICF, publicized in the early 1960s, consists of targeting a small fuel sphere with high-energy laser beams (``direct-drive'') or X-rays (in the case of ``indirect drive'') resulting from the conversion of the energy of laser beams focused inside a millimeter-size gold cavity or Hohlraum \cite{Atzeni2004}. The outer layer of the target is suddenly transformed into a plasma and the fuel (a mix of deuterium and tritium) is quickly compressed by the rocket-like blow-off. When the fuel reaches high density, thermonuclear burn happens within a nanosecond, before the fuel can bounce and spread. As a result, its own inertia is responsible for the confinement. The fusion of deuterium and tritium cores produces $\alpha$-particles (He$^{2+}$) which deliver energy to the plasma and thus heat the fuel, allowing more cores to fuse. This implies that the stopping power of $\alpha$-particles in the plasma is a key parameter of ICF. Initially, ICF was explored using laser pulses. In the 1970s, the development of intense ion beams (Humphries {\it et al.} 1974 \cite{Humphries1974}) was followed by their application to ICF (Prono {\it et al.} 1975 \cite{Prono1975}, Poukey 1975 \cite{Poukey1975}): ion beams were then more powerful than laser beams and reached the 50 TW/cm$^2$ required for ICF \cite{ICF}. As a result, plasma stopping power for ions became a significant field of research. 

In the medical field, the use of protons for radiotherapy of cancer was suggested in the 1940s. Van Abbema {\it et al.} \cite{Protontherapy} recently emphasized that proton beams have a finite range and local high dose region, which means that it is possible to precisely target the tumor without exposing healthy tissues. As a consequence, protons could replace photons and electrons, provided the proton stopping powers of human tissues are known with a reasonable level of accuracy. 

From a broader perspective, stopping powers are commonly used in radiation protection as a way to estimate the degree of danger caused by the exposure to radiation. They also find industrial applications, such as ion implantation for semiconductor devices. Stopping powers are key elements in the study of high-energy-density matter, for instance in order to measure equations of state of isotropically expanded warm plasmas created by isochoric heating.

The ability to perform fast calculations of stopping powers is of great interest, for instance in the context of radiative-hydrodynamics simulations of Hohlraums in the framework of the design and interpretation of laser-fusion experiments. Indeed, a realistic modeling requires inline calculations of opacity and emissivity at each time step and in each spatial cell of the radiative-hydrodynamics simulation. The computation time is therefore the main limiting factor, and developing fast approximate (i.e., as much analytical as possible) methods for the calculation of stopping powers becomes an important challenge.

The present work was stimulated by the 2016 review of Deutsch and Maynard \cite{Deutsch2016} about ion stopping in dense plasma. The main goal of the present work is to investigate parametric potentials to compare their predictions to recent average-atom codes. In the warm-dense-matter (WDM) regime, we use an analytical approach as often as possible to develop a simple model, which is more flexible and explicit than self-consistent mean-field average-atom codes and leads to fast approximations of stopping powers for projectiles with energies between 0.3 and 100 MeV. We focus on protons and $\alpha$-particles traveling in an aluminum (Al) plasma with different degrees of ionization.

In the present work, we focus on the plasma target, the main purpose of our work consisting in taking into account its partial ionization, since most of the stopping-power models available in the literature are restricted to neutral or fully-ionized plasmas. This is the reason why we consider mid-$Z$ ($Z$ being the atomic number) elements. Although it is definitely true that aluminum and iron are not key elements in ICF, they are still representative. Indeed, in ICF, the $ \alpha$ particles are slowed down in mid-$Z$ plasmas as well, such as plastic (CH) for the ablator, silicon or germanium, used as dopants in the ablator, and possibly gold from the Hohlraum \cite{Poujade2015}.

The main assumptions of the formalism are described in section \ref{sec1}, with a particular emphasis on the plasma mean excitation energy. The parametric potentials are presented in section \ref{sec2}, together with the average quantities required for the calculation of the stopping power. Comparisons with the self-consistent average-atom code {\sc atoMEC} are discussed in section \ref{sec3}, and prescriptions for the calculation of the total stopping power for neutral, fully-ionized and partially-ionized plasmas are given in section \ref{sec4}.

\section{Stopping power of bound electrons}\label{sec1}

The ion stopping power, $S$, is due to scattering events in the plasma. It can be calculated considering three contributions related to collisions with the plasma ions (I), the electrons bound (B) to the ions and the free (F) electrons. As the nuclear term is rather negligible compared to the others, we concentrate on the free and bound electrons. In that framework, the stopping power reads
\begin{equation}\label{threecom}
    S = -\frac{dE}{dx}= -\frac{dE}{dx}\Bigl|_{\bf I} -\frac{dE}{dx}\Bigl|_{\bf B} -\frac{dE}{dx}\Bigl|_{\bf F} \approx -\frac{dE}{dx}\Bigl|_{\bf B} -\frac{dE}{dx}\Bigl|_{\bf F},
\end{equation}
where $E$ represents the energy of the projectile and $x$ the plasma penetration length.

\subsection{Bound electrons stopping power: the Bethe-Bloch formula}

In 1930, using quantum differential cross sections calculations \cite{Bethe1930}, Bethe expressed the stopping power of bound electrons using a perturbation-theory expansion as
\begin{equation}\label{bethe}
    -\frac{dE}{dx}\Bigl|_{\bf B}=\frac{4\pi \left[Z_{\mathrm{eff}}(V_p)~e^2\right]}{m_eV_p^2}n_T \, L_e,
\end{equation}
where $Z_\mathrm{eff}$ is the effective charge of the projectile, $V_p$ its velocity, $n_T= N_A Z_T \rho_{\mathrm{i}} /A_T$ the electron density of the target, $Z_T$ its atomic number and $A_T$ its mass number. $N_A$ represents the Avogadro number and $\rho_{\mathrm{i}}$ the (mass) density of the material. Finally, $L_e$ is the stopping number, expressed by Bethe as $\ln\left(2m_eV_p/I\right)$, $I$ being the mean excitation energy of the ion and $m_e$ the electron mass. Most of our efforts are dedicated to the evaluation of the quantity $I$ using parametric potentials as described by Deutsch and Maynard \cite{Deutsch2016}. The Bethe formula (\ref{bethe}) remains a good starting point for the determination of the stopping power of bound electrons, especially because many corrections have been added to the original formula since 1930. In that framework, the stopping number can be put in the form $L_e = L_0 + ZL_1 + Z^2L_2$. In order to take into account the relativistic effects at high projectile velocities, the main term $L_0 = \ln\left(2m_eV_p/I\right)$ was replaced by Bethe in 1932 by
\begin{equation*}
    L_0 = \ln\left[\frac{2m_e \beta^2 c^2}{I \left(1-\beta^2\right)}\right] - \beta^2 \qquad \mathrm{with} \qquad \beta = \frac{V_p}{c},
\end{equation*}
where $c$ denotes the speed of light. The term $L_0$ can be completed with two more corrections $L_1$ and $L_2$ (see Appendix \ref{appA}). 

When decelerated to velocities close to those of the bound electrons, incident ions can capture electrons. As a result, the Bethe formula assumes an equilibrium effective charge (the mean charge of the ion beam) $Z_{\mathrm{eff}}$ instead of $Z_p$, for the projectile. Following the Thomas-Fermi model, Northcliffe \cite{Northcliffe1960} showed that 
\begin{equation*}
    \frac{Z_{\mathrm{eff}}}{Z} = 1 - \exp\left[-\frac{V_p}{v_0\,Z^{2/3}}\right],
\end{equation*}
$v_0$ being the orbital velocity of the electron being captured (or lost). Using this frame, different semi-empirical models have been developed, such as the Betz formula \cite{Betz1970} or the following form suggested by Brown and Moak \cite{Brown1972}:
\begin{equation}\label{bro}
    \frac{Z_{\mathrm{eff}}}{Z} \approx 1 - 1.034 \cdot \exp\left[-137.04\,\frac{\beta}{Z^{0.69}}\right].
\end{equation}
The coefficients of such models are determined by measuring stopping powers for different incident heavy ions passing through the same target. Both the incident and resulting velocities of the projectiles are measured for a thin layer of material, such that the ionic velocity can be considered as almost constant during the travel in the material. As the Bethe formula in its simplest form can be written (forgetting the dependence on the target) $S = Z_{\mathrm{eff}}^2~f(V_p)$, for ions with the same velocity, the stopping powers differ by their effective charge only. This method might lead to significant bias as the effective charges are deduced from the knowledge of some measured stopping powers, in order predict unknown ones. Moreover, the expressions of $Z_{\mathrm{eff}}$ do not include time-dependency and the average value is certainly not instantaneously reached. Traveling at a constant speed, an ion beam effective charge sees its charge progressively change. Finally, taking the high velocity limit of Brown and Moak's expression leads to $Z_{\mathrm{eff}}=Z$, which makes sense as the projectile does not have time to interact with the surrounding matter. However, in the limit of low velocities, nonphysical negative values of $Z_{\mathrm{eff}}$ can be encountered. 

As a result of the previously listed limits, the semi-empirical methods can only be used for high energies heavy ions and should be carefully considered, even if Gauthier {\it et al.} \cite{Gauthier2013_bis} showed that these models give comparable values for heavy ions. In the present work, since the incoming beam is exclusively made of protons, Gus'kov {\it et al.} \cite{Gus'kov2009} suggest that there is no need to consider an effective charge. We make the same assumption for $\alpha$-particles, since our main interest lies in the evaluation of $I$.

\subsection{The mean excitation energy $I$ \label{I_Explained}}

Now that all the correction terms for the Bethe formula have been defined, the mean excitation energy, $I$, must be calculated. By definition, it can be expressed as
\begin{equation*}
    \ln(I) = \frac{\sum_n f_{0n} \ln(E_{0n})}{\sum_n f_{0n}},
\end{equation*}
with $f_{0n}$ the optical oscillator strengths of a transition of the ion's electrons from the fundamental state $\ket{0}$ to the excited state $\ket{n}$ in a scattering event. $E_{0n}$ is the energy difference between the states 0 and $n$. With $\vec{r}_i$ the position of the $N$ bound electrons and ${\bf q}$ the impulsion transferred, one can write
\begin{eqnarray*}
    f_{0n} = \lim_{q\to 0} \quad \frac{E_{0n}}{NQ} \lvert \bra{n} \sum_{j=1}^N e^{i {\bf q \cdot r_i}} \ket{0} \rvert ^2 \quad \mathrm{with}\quad Q = \frac{\hbar^2 q^2}{2m_e}.
\end{eqnarray*}
As a result, $\ln(I)$ corresponds to the normalized sum of excitation energies, weighted by the transition probabilities. Hence, the mean excitation energy represents the mean energy transmitted by the projectile to the target ions in every collisional event. In order to calculate oscillator strengths, one needs to carry out complex and time-consuming quantum computations. As our purpose is to find analytical formulas involving the use of parametric potentials, we follow the formalism employed by Garbet \cite{Garbet1984}. Let us first define the sum rules
\begin{equation*}
    S(\mu) = \sum_n f_{0n} E_{0n}^{\mu}
\end{equation*}
and
\begin{equation*}
    L(\mu) = \frac{\,dS(\mu)}{\,d\mu} = \sum_n f_{0n} E_{0n}^\mu \ln\lvert E_{0n} \rvert,
\end{equation*}
where $I$ can be expressed in the form:
\begin{equation*}
    \ln(I) = \frac{L (0)} {S (0)} = \frac{d}{d\mu} \ln S(\mu) \big\rvert_{\mu = 0}.
\end{equation*}
Even though $L(0)$ cannot be easily calculated, it is possible to find an approximation for $I$ using a rather simple method. For a subshell $\alpha$ and considering subshells $\alpha '$ such as $E_{\alpha'} > E_\alpha$, alternative sum rules can be considered, in particular
\begin{equation*}
    S_\alpha (\mu) = \sum_{\alpha < \alpha'} (1-h_{\alpha '}) f_{\alpha \alpha'} E^\mu_{\alpha \alpha '}
\end{equation*}
and
\begin{equation*}
    L_\alpha (\mu) = \sum_{\alpha < \alpha'} (1-h_{\alpha '}) f_{\alpha \alpha'} E^\mu_{\alpha \alpha '} \ln \lvert E_{\alpha \alpha '} \rvert,
\end{equation*}
where $h_{\alpha '}$ is the number of electrons of the subshell $\alpha '$ divided by its degeneracy. It corresponds actually to an occupation rate and accounts for the transitions blocked by electrons already in $\alpha '$. Due to the concavity of the logarithm, for $F_k$ such that $F_k > 0$ and $\sum_k F_k =1$, the Jensen inequality enables one to deduce that
\begin{equation*} 
    \sum_k F_k \ln(E_k) \leq \ln \left(\sum_k F_k E_k\right) 
\end{equation*}
as well as
\begin{equation*} 
    -\ln \left(\sum_k \frac{F_k}{E_k}\right) \leq -\sum_k F_k \ln(E_k),
\end{equation*}
and taking $E_k = E_{\alpha \alpha'}$ and $F_k = F_{\alpha \alpha '} = (1-h_{\alpha '}) f_{\alpha \alpha'} E^\mu_{\alpha \alpha '} / S_\alpha (\mu)$, we can write
\begin{equation*}
    \ln \left[\frac{S_\alpha(\mu)}{S_\alpha(\mu -1)}\right] \le \frac{L_\alpha(\mu)}{S_\alpha(\mu)} \le \ln \left[\frac{S_\alpha(\mu +1)}{S_\alpha(\mu)}\right].
\end{equation*}
For an atom or an ion, the same inequality is verified by the total sum rules $S^P (\mu) = \sum_\alpha g_\alpha S_\alpha (\mu)$ and $L^P (\mu) = \sum_\alpha g_\alpha L_\alpha (\mu)$. The $P$ stands for ``physical'' as these rules imply $g_\alpha$, the number of electrons in the subshell $\alpha$. Garbet also showed that the physical sum rules verify $S^P (1) < S (1)$, $S^P (-1) < S(-1)$ and $S^P (0) = S(0)$. As a result, taking $\mu=0$, one has
\begin{equation}\label{ln(I)}
    \ln \left[\frac{S(0)}{S(-1)}\right] \le \frac{L(0)}{S(0)} = \ln(I) \le \ln \left[\frac{S(1)}{S(0)}\right]
\end{equation}
and $\ln(I)$ can be approximated by the half-sum of the lower and upper bounds of the above inequality:
\begin{equation*}
    \frac{1}{2}\left\{\ln \left[\frac{S(0)}{S(-1)}\right]+\ln \left[\frac{S(1)}{S(0)}\right]\right\}=\frac{1}{2}\ln \left[\frac{S(1)}{S(-1)}\right].
\end{equation*}

We develop below the method to reach $S(1)$ and $S(-1)$. Let us write the quantum operators as $\hat{p}_{zi}$ and $\hat{z}_i$ for the impulsion and position of the $i^{th}$ electron. If the potential $\hat{V}$ accounts for all core-electron and electron-electron interactions, the Hamiltonian of the system can be written $\hat{H}=\sum_{i} \hat{p}_{zi}^2/2m_e + V(r_i)$. In this case, the following commutators can be calculated:
\begin{equation}\label{Commutator1}
    [\hat{H}-E_0, \hat{z_i}] = [\frac{\hat{p}_{zi}^2}{2m_e}, \hat{z}_i] = -i\hbar \frac{\hat{p}_{zi}}{m_e},
\end{equation}
and
\begin{align}\label{Commutator2}
    [\hat{H}-E_0, \hat{p}_{zi}]\hat{\Psi} &= [\hat{V}, \hat{p}_{zi}]\Psi\nonumber\\
    &= -i\hbar \hat{V} \cdot \frac{\partial\Psi}{\partial z_i} + i\hbar \frac{\partial}{\partial z_i} (\hat{V}\Psi)\nonumber\\
    &= i\hbar \frac{\partial V}{\partial z_i} \cdot \Psi.
\end{align}
Consequently, the oscillator strengths can be re-formulated as follows. First, setting $z = {\bf q \cdot r}/q$, one finds
\begin{equation}\label{fon1}
    f_{0n} = \frac{2m_eE_{0n}}{\hbar^2N} \lvert \langle n| \sum_{i=1}^N z_i |0\rangle \rvert ^2.
\end{equation}
As $(\hat{H}-E_0) \ket{n} = E_{0n} \ket{n}$, one gets
\begin{align*}
     f_{0n} &= \frac{2m_e}{\hbar^2N}\bra{n} \sum_{i=1}^N z_i \ket{0} \bra{0} \sum_{j=1}^N z_j (\hat{H}-E_0) \ket{n}\nonumber\\
     &= \frac{2m_e}{\hbar^2N}\bra{n} \sum_{i=1}^N z_i \ket{0} \bra{0} \sum_{j=1}^N \left[-i\hbar \frac{\hat{p}_{zj}}{m_e} + (\hat{H}-E_0)z_i\right] \ket{n},
\end{align*}
where we used the commutator (\ref{Commutator1}). Finally, as $\bra{0} (\hat{H}-E_0) = \bra{0}(E_0-E_0)$, one finds
\begin{equation}\label{fon2}
    f_{0n} = -\frac{2i}{\hbar N}\bra{n} \sum_{i=1}^N \hat{z_i} \ket{0} \bra{0} \sum_{j=1}^N \hat{p}_{zj} \ket{n},
\end{equation}
and with the commutator (\ref{Commutator2}), one obtains 
\begin{equation}\label{fon3}
    f_{0n} = \frac{2}{E_{0n}N}\bra{n} \sum_{i=1}^N \hat{z_i} \ket{0} \bra{0} \sum_{j=1}^N \frac{\partial V}{\partial z_j} \ket{n}.
\end{equation}
These forms of the oscillator strengths make possible the evaluation of the sum rules. For $S(-1)$, taking Eq. (\ref{fon1}) and using the closure relation,
\begin{align}\label{S(-1)}
    S(-1) &= \frac{2m_e}{\hbar^2N} \sum_{n \ne 0} \lvert \bra{n} \sum_{i=1}^N \hat{z_i} \ket{0} \rvert ^2\nonumber\\
    &= \frac{2m_e}{\hbar^2N} \sum_{n \ne 0} (\frac{1}{3}\bra{0} r^2 \ket{0} + \bra{0} \sum_{i \ne j} z_i z_j \ket{0}).
\end{align}
We used the fact that, on average, $\langle z^2\rangle = \langle r^2\rangle/3$. Thus, neglecting correlation terms, one gets
\begin{equation*}
    S(-1) = \frac{2m_e}{3\hbar^2} r_0^2
\end{equation*}
with $r_0^2 = \langle r^2\rangle$, the mean squared radius. In the same way, using Eq. (\ref{fon3}) for $S(1)$ and neglecting correlation terms, 
\begin{equation}\label{S(1)}
    S(1) = \frac{4}{3N} \bra{0} \frac{\hat{p}^2}{2m_e} \ket{0} = \frac{4}{3} K_0^2
\end{equation}
with $K_0$ the mean kinetic energy. Combining the two preceding equations to the mean value of the inequality (\ref{ln(I)}), the final expression for $I$ is:
\begin{equation}\label{Calcul_I}
    I = \frac{e^2}{a_0} \sqrt{\displaystyle\frac{2K_0 a_0}{e^2 \left\langle\displaystyle\frac{r^2}{a_0^2}\right\rangle}} = \sqrt{\frac{2K_0}{r_0^2}}.
\end{equation}
Consequently, we need to find the kinetic energy and the mean squared radius for ions in a plasma. In the following this method of calculation of the mean excitation energy $I$ will be referred to as Garbet's method. In order to keep an analytical approach, effective potentials will be used to represent average atoms. However, before working on potentials, let us consider the uncertainties related to such an approximation. Two main assumptions are made. First, correlation terms of $S(1)$ and $S(-1)$ (Eq. (\ref{S(1)}) and (\ref{S(-1)})) are negligible. Then, $\ln(I)$ is close to the mean of inequality (\ref{ln(I)}). Estimating the influence of these approximations (or overcoming them) would require calculating the oscillator strengths. Nevertheless, it is possible to estimate the total error with the work of Dehmer {\it et al.} \cite{Dehmer1975} who used an Hartree-Fock method to determine the wave functions and energies of the electrons. They give values for the different sum rules for atoms. Considering all atoms from $Z=2$ to $Z=18$ and assuming that the percentage error on the expected value $L(0)/S(0)$ is estimated as $0.5 \cdot \ln (S(1) / S(-1))$, the error was found to be between $2$ and 12 \% (for Ne). However, the error for most atoms is below 8 \%, which is reasonable considering that the values of $I$ found by Dehmer {\it et al.} are sometimes far from the experimental values they gathered. There is for instance a 24 \% difference for Al, leading to a 5.3 \% difference on $\ln(I)$. Nevertheless, Garbet's method provides a good insight into the behavior of $I$ and an estimate of the error, provided that $S(0)$ is known. In the work of Dehmer {\it et al.}, the definition of $S(0)$ does not coincide with Garbet's one. As a result, it cannot be used to estimate the gap between $\ln \left[S(0)/S(-1)\right]$ and $\ln \left[S(1)/S(0)\right]$. In order to get an idea of the uncertainty of our results, we consider that $S(0) = \sum_n f_{0n}$ should be normalized consistently with the Thomas-Kuhn-Reiche sum rule, such that the occupation number of 1s for the average-atom is 2. Thus, we have $S(0) = 2$. (Garbet suggests $S(0)=1$ \cite{Garbet1984}).

\section{Analytical Potentials}\label{sec2}

For a sake of simplicity and since we are interested in the most analytical model as possible for ion stopping, we resort to parametric potentials to represent the mean ion of the plasma and follow the formalism developed in the work of Garbet \cite{Deutsch2016, Garbet1984} to determine the mean excitation energy. All the considered parametric potentials have a similar shape: each one of the $N$ electrons is subject to the potential. We write
\begin{equation*}
    V(r)=-\frac{(Z_T-N+1)}{r}+\phi(r)
\end{equation*}
with 
\begin{equation*}
    \phi(r)=-\frac{N}{r}\Omega(r).
\end{equation*}
The potential $V(r)$ is the sum of the Coulomb potential due to the nucleus and the $N-1$ remaining electrons if they were all at the same distance from the nucleus. $\Omega$ is the screening function, taking into consideration the spatial scattering of the electrons. Using such a potential, it is possible to find an expression for the mean excitation energy. 

\subsection{From potentials to mean excitation energies}

As previously stated in Eq. (\ref{Calcul_I}), 
\begin{equation*}
    \langle I\rangle=\sqrt{\frac{2K_0}{r_0^2}}.
\end{equation*}
The calculation of $r_0^2$ requires knowing the density, which can be found using the Poisson equation:
\begin{equation}\label{Poisson}
    \Delta\phi(r)=- 4\pi\rho(r) =\frac{1}{r^2}\frac{\partial}{\partial r}\left[r^2\frac{\partial}{\partial r}\phi(r)\right].
\end{equation}
As the total number of bound electrons is preserved, since
\begin{equation*}
    \int_0^{\infty}4\pi r^2\rho(r)\,\mathrm{d}r=N,
\end{equation*}
the mean square radius is
\begin{equation}\label{R02}
    r_0^2=\langle r^2\rangle=\frac{\displaystyle\int_0^{\infty}r^24\pi r^2\rho(r)\,\mathrm{d}r}{\displaystyle\int_0^{\infty}4\pi r^2\rho(r)\,\mathrm{d}r}.
\end{equation}
For the kinetic energy $K_0$, Garbet suggested to apply the Virial theorem:
\begin{equation}\label{Viriel}
    2K_0=-\frac{\langle V(r)\rangle}{N}
\end{equation}
yielding
\begin{equation}\label{K0}
    2K_0=-\frac{1}{N}\int_0^{\infty}\left\{-\frac{Z_T}{r}+\frac{N}{2r}\left[1-\Omega(r)\right]\right\}4\pi r^2\rho(r)\,\mathrm{d}r.
\end{equation}
The factor $1/2$ in front of the electron-electron interaction term $N\left[1-\Omega(r)\right]/r$ avoids counting the contribution of each electron twice (this does not perfectly correspond to the formula suggested in Ref. \cite{Deutsch2016}). We have considered different forms for the screening function $\Omega(r)$, which are detailed below. 

\subsection{The GSZ potential}

The two-parameters potential of Green, Sellin and Zachor (GSZ) \cite{Green1969} reads
\begin{equation*}
    \Omega_{\mathrm{GSZ}}(r)=\frac{1}{H(e^{r/d}-1)+1}.
\end{equation*}
The parameters $H$ and $d$ have to be determined for each $N$ and $Z_T$. Using the Poisson equation (\ref{Poisson}) and Eq. (\ref{R02}), we obtain 
\begin{equation*}
    Nr_0^2=-6\int_0^{\infty}r^2\phi(r)\,\mathrm{d}r.
\end{equation*}
After integrating by parts and setting $r_G^2=6d^2N/H$, one finds
\begin{equation*}
    Nr_0^2=r_G^2F(\alpha)
\end{equation*}
with $\alpha=1-1/H$, and
\begin{equation*}
    F(\alpha)=\sum_{n=0}^{\infty}\frac{\alpha^n}{(n+1)^2}=\frac{1}{a}\mathrm{Li}_2(a),
\end{equation*}
where $\mathrm{Li}_n$ is the usual polylogarithm function:
\begin{equation*}
    \mathrm{Li}_n(z)=\sum_{k=1}^{\infty}\frac{z^k}{k^n}.
\end{equation*}
The calculation of moments of the electron density can be generalized as
\begin{equation*}
    \langle r^n\rangle=\frac{\displaystyle\int_0^{\infty}r^n4\pi r^2\rho(r)\,\mathrm{d}r}{\displaystyle\int_0^{\infty}4\pi r^2\rho(r)\,\mathrm{d}r}
\end{equation*}
and one gets
\begin{equation*}
    N \langle r^n\rangle=\frac{(n+1)!d^n}{H-1}~\mathrm{Li}_n\left(\displaystyle\frac{H-1}{H}\right).
\end{equation*}
Then, using equation (\ref{K0}), one gets
\begin{equation*}
    2K_0=-\frac{N+2HN-12HZ_T}{12d}=\frac{H}{d}\left[Z_T-N\left(\frac{1}{6}+\frac{1}{12H}\right)\right],
\end{equation*}
and finally (note that Green {\it et al.} suggest, in order to simplify the expressions, that $\frac{H}{d}\approx Z_T^{\eta}$ yielding $r_G^2\approx 6N^{1-\eta}d$ with $\eta=0.4$ \cite{Green1969} but we chose not to use this approximation):
\begin{equation*}
    \langle I\rangle^2=\frac{1}{6}\left(\frac{H}{d}\right)^2\frac{Z_T}{d}\left[1-\frac{N}{6Z_T}\left(1+\frac{1}{2H}\right)\right]\displaystyle\frac{1}{F\left(1-\displaystyle\frac{1}{H}\right)}.
\end{equation*}
A set of parameters for the GSZ potential has been suggested by Garvey, Jackman and Green (GJG) \cite{Garvey1975} using a minimization of energy based on a modified Hartree-Fock method. GJG point out that parameters $1/d$ and $H/d$ for a fixed $N$ exhibit a linear variation with respect to $Z_T$. As a result, for each $N$, they give linear fits of these parameters as functions of $Z_T$. 

\subsection{The Yunta potential}

Other potentials have been suggested by the literature. An interesting example is the Yunta potential \cite{Yunta1974} defined by
\begin{equation*}
    \Omega_Y(r)=(1-c\,r)\exp\left(-a\,r^f\right).
\end{equation*}
The Poisson equation (\ref{Poisson}) gives 
\begin{equation*}
    4\pi r^2\rho(r)=-afNr^{f-1}\,e^{-a\,r^f}\left[f(c\,r-1)(a\,r^f-1)-1-c\,r\right]
\end{equation*}
still satisfying
\begin{equation*}
    \int_0^{\infty}4\pi r^2\rho(r)\,\mathrm{d}r=N,
\end{equation*}
and one finds
\begin{eqnarray*}
    Nr_0^2=\frac{6N}{f} a^{-\frac{3}{f}} \left[a^{\frac{1}{f}}~ \Gamma\left(\frac{2}{f}\right) - c~ \Gamma\left(\frac{3}{f}\right)\right],
\end{eqnarray*}
where $\Gamma$ is the usual Gamma function. For the determination of $K_0$, there are some constraints for the different parameters. We have indeed to compute
\begin{align}\label{K0Y}
    &\int_0^{\infty}r^{f-2}\,e^{-2a\,r^f}\left\{\vphantom{\frac{a}{a}}f(c\,r-1)(a\,r^f-1)\right.\nonumber\\
    &\;\;\;\;\;\;\left.\left[N(cr-1)+e^{a\,r^f}(N-2Z_T)\right]-1-c\,r\right\}\,\mathrm{d}r
\end{align}
which means that we need to calculate integrals of the type
\begin{equation*}
    \int_0^{\infty}r^{\mu}~e^{-ar^f}\,\mathrm{d}r
\end{equation*}
which converge only if $\mu>-1$ and $f>0$. In Eq. (\ref{K0Y}), we encounter particular integrals such as
\begin{equation*}
    \int_0^{\infty}r^{f-2}~e^{-br^f}\,\mathrm{d}r,
\end{equation*}
which requires $f-2>-1$, {\it i.e.}, $f>1$. This means that the Yunta-Martel potential is not compatible with the Virial theorem (in the Garbet application \cite{Garbet1984}), since the average energy is not defined. Consequently, we obtain $K_0$ by numerically integrating with Simpson's second rule. 

It is worth mentioning that Martel {\it et al.} \cite{Martel1995} used the Nelder-Mead non-linear simplex method \cite{Nelder1965} in order to minimize the total energy of the ion and determine the parameters $a$, $c$ and $f$. From the general shape of the potential, some conditions for the parameters can be found. For instance, it is obvious that if $f>0$, $a$ must be greater than $0$, so that $\lim_{r\to\infty} V(r) = 0$. We noticed that these conditions are not always satisfied by Martel's parameters, resulting in a few values of ($N, Z_T$) for which the parameters ($a, f, c$) need to be corrected. 

\subsection{The Klapisch potential}

The Klapisch potential \cite{Klapisch1971,Tannous1999} is widely used in atomic-structure calculations. It reads
\begin{equation*}
    \Omega_K(r)=e^{-\alpha_1 r}+\frac{C}{N}\,r\,e^{-\alpha_2r}.
\end{equation*}
The corresponding Poisson equation yields
\begin{equation*}
    4\pi r^2\rho(r)=r\,e^{-(\alpha_1+\alpha_2)\,r}\left[N\alpha_1^2e^{\alpha_2\,r}+C\alpha_2\left(\alpha_2\,r-2)e^{\alpha_1\,r}\right)\right]
\end{equation*}
still satisfying
\begin{equation*}
    \int_0^{\infty}4\pi r^2\rho(r)\,\mathrm{d}r=N,
\end{equation*}
and leading to
\begin{equation*}
    Nr_0^2=\int_0^{\infty}4\pi r^4\rho(r)\,\mathrm{d}r=\frac{6N}{\alpha_1^2}+\frac{12C}{\alpha_2^3}.
\end{equation*}
More, generally, the $n^{th}-$order moment of the electron density reads
\begin{equation*}
    N\langle r^n\rangle=\frac{(n+1)!N}{\alpha_1^n}+\frac{3^{n-1}\times 2^nC}{\alpha_2^{n+1}}.
\end{equation*}
In the same way, one gets
\begin{align*}
    2K_0=&-\frac{1}{N}\int_0^{\infty}\left\{-\frac{Z_T}{r}+\frac{N}{2r}\left[1-\Omega(r)\right]\right\}4\pi r^2\rho(r)\,\mathrm{d}r\\
    =&\left(Z_T-\frac{N}{4}\right)\alpha_1+\frac{C\alpha_1^2}{(\alpha_1+\alpha_2)^2}-\frac{C(C+8Z_T\alpha_2)}{8N\alpha_2}.
\end{align*}

In the literature, no universal parameters have been defined for the Klapisch potential. Since the development of an energy-minimizsation code was beyond the scope of the present work, we tried to determine them using the other potentials investigated in this work. As Martel's parameters are not complete and as the Yunta potential does not lead to an analytical formula for $K_0$, we decide to consider the GSZ potential as a reference. The determination of the parameters of the Klapisch potential is discussed in Appendix \ref{appB}.

\subsection{Quantum shell effects in parametric potentials}

\begin{figure*}[ht!]
    \centering
        \includegraphics[scale=0.35]{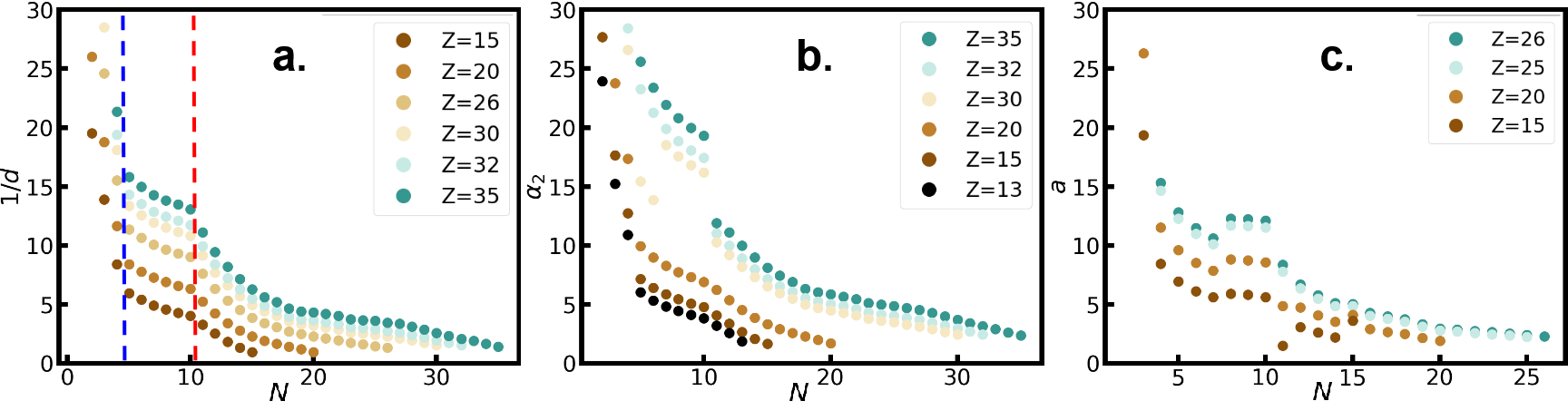}
    \caption{{\bf (a):} Parameter $1/d$ of the GSZ potential as defined by GJG for different elements, as a function of the number of bound electrons, $N$. Vertical lines: transition from 4 to 5 electrons and from 10 to 11.
    {\bf (b):} Same graph for the parameter $\alpha_2$ of the Klapisch potential as defined by our previous work. {\bf (c):} Parameter $a$ of the Yunta potential, as defined by Martel. \label{Parametres}}
\end{figure*}

As illustrated by Fig. \ref{Parametres}, the GSZ parameter $1/D$ as defined by GJG reflects the subshell structure. What is remarkable, is that the Klapisch potential seems to increase the gaps between different shells. It might come from the general shape of the potential including one more parameter. The quantum effects on the Klapisch parameters come from similar effects as for the GSZ potential . On the contrary, the Yunta potential does not display as obvious transitions. The transition from 2p to 3s is still very clear. The evidence for the signature of the quantum effects represents one more advantage of the GSZ potential with GJG parameters over the Yunta potential.

\section{Comparison of the parametric potentials to average atom model code {\sc atoMEC}}\label{sec3}

\subsection{Electron radial density}

It seems interesting to try to compare the different parametric potentials to a reference potential defined by an average-atom code. We use {\sc atoMEC} (average-ATOm code for studying Matter under Extreme Conditions) developed by Callow {\it et al.} \cite{Callow2022}, as it is to our knowledge the only open-source code and it is developed in Python. In the warm dense matter regime, the quantum distribution of the electrons matters. However, Callow {\it et al.} pointed out that the calculation cost in the Density Functional Theory (DFT) scales as {\it O}($N^3 \tau^3$) where $\tau$ is the temperature. This is why {\sc atoMEC} considers an average-spherical-atom model: it compares with the parametric potentials previously defined. The potential is calculated with the Kohn-Sham DFT (KS-DFT). First, the plasma and the atom model parameters are defined. These parameters are, for instance, the ($n,\ell$) subshells ($n$ and $\ell$ being respectively the principal and orbital quantum numbers) and their associated wavefunctions (with their buondary conditions) and eigen-energies that will be considered by the code as accessible for the electrons of the average-atom. Then, a guess for the radial KS-orbitals and the Fermi-Dirac distribution enables one to construct the density and the potential. The KS equation (the Schr\"odinger equation for a non-interaction system) is solved to calculate the orbitals. The loop is repeated until the energy, density and potential converge. This is a self-consistent-field (SCF) procedure. In particular, {\sc atoMEC} provides the radial density. 

\begin{figure*}[ht!]
    \centering
        \includegraphics[scale=0.40]{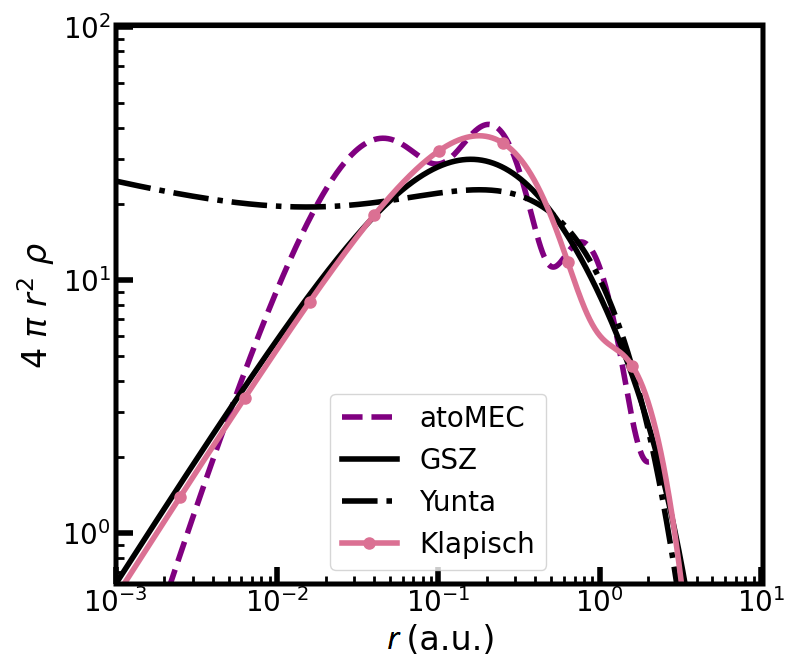}
    \caption{Comparison of the radial density for Fe obtained using the Poisson equation with the GSZ potential using GJG parameters. \label{Density}}
\end{figure*}

Fig. \ref{Density} displays the normalized radial density of Fe as obtained with {\sc atoMEC} or the parametric potentials. A first interesting point is that all densities show similar behaviors at high $r$, even though the Yunta radial density does not go to zero for the small values of $r$ while the {\sc atoMEC}, GSZ or Klapisch ones do. Also, the Klapisch density leads to interesting oscillations, which remind those of the {\sc atoMEC} code, related to the quantum distribution of electrons. It could be interesting to define the Klapisch parameters by minimizing the ionic energy to see it the density fits the {\sc atoMEC} one. 

\begin{figure*}[ht!]
    \centering
        \includegraphics[scale=0.32]{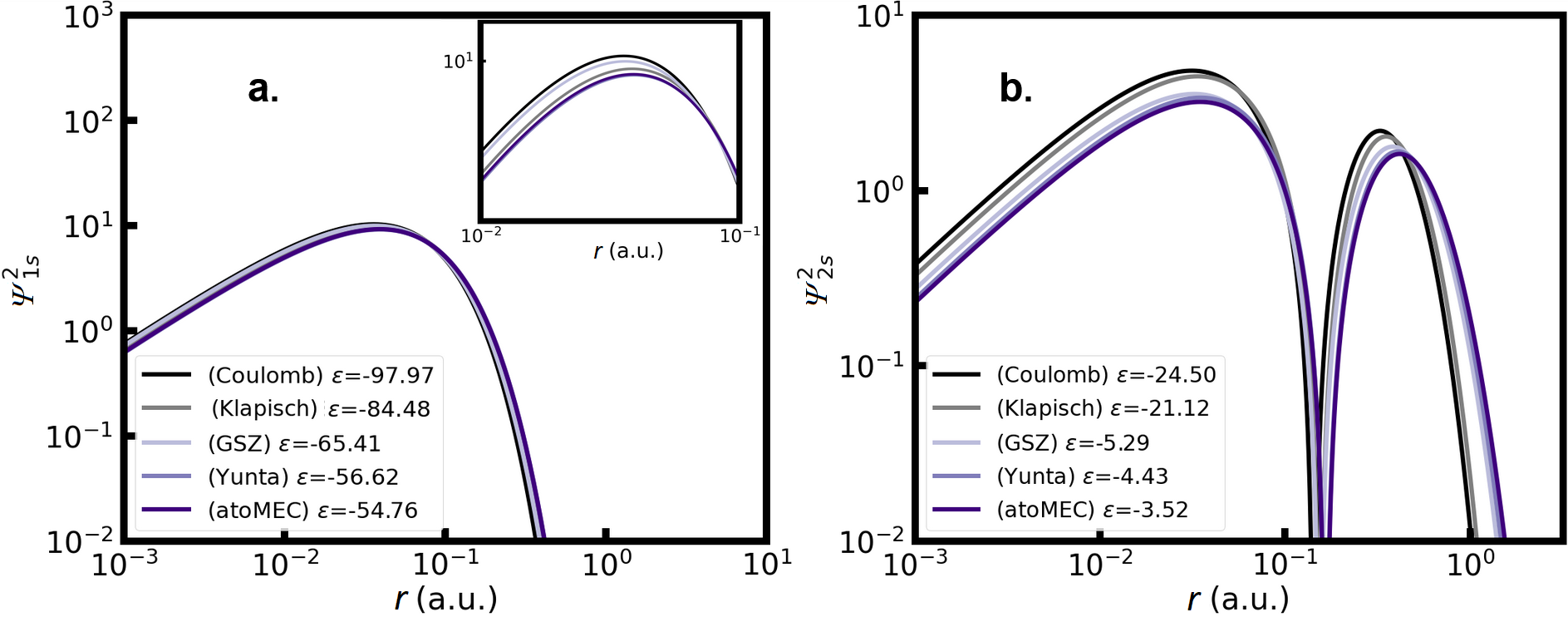}
    \caption{Squared wave functions defined by {\sc atoMEC} using different potentials: Klapisch, GSZ, Yunta, but also a Coulomb potential and the optimal potential defined by {\sc atoMEC}. The eigenvalues $\epsilon$ related to these orbitals are given in a.u. {\bf (a):} Orbital 1s. {\bf (b):} Same graph for the orbital 2s. \label{Orbitals}}
\end{figure*}

For the calculation of $I$, the key parameter is not the density but the one-electron energies. Using {\sc atoMEC} and changing the initial guess for the radial potential, the Schr\"odinger equation is solved (using a matrix implementation of the Numerov scheme) to find the radial KS-orbitals for the parametric potentials and the related energies $\epsilon_{n\ell}$. Figure \ref{Orbitals} shows the squared wavefunctions of subshells 1s and 2s obtained with the parametric potentials and stemming from the SCF procedure of {\sc atoMEC}. The Yunta potential leads to the lowest energies of all potentials and is really close to the {\sc atoMEC} result. On the contrary, the Klapisch potential, as expected with the way its parameters are defined, gives less accurate energies, still better than the Coulomb potential however. 

Getting the eigenvalues allows us to calculate, for the parametric potentials, the kinetic energy of the electrons and to compare it to the the mean kinetic energy found using the Virial theorem, as suggested by Garbet. The error done using the Virial theorem seems to be systematically around 1 to 2\% for Al, Si and Fe. As a result, the use of the Virial theorem, even if it is not perfectly justified in this system, does not lead to a significant error on the mean excitation energy. \\

As a conclusion, so far we have used four different potentials, two of them having parameters defined by energy minimization. The GSZ and the Klapisch potentials lead to the most interesting radial densities and their parameters seem to be shaped by quantum effects. However, the Yunta potential leads to the lowest energies and seems to be the best candidate to calculate mean excitation energies and the stopping power.

\subsection{Values of the mean excitation energy $I$ in plasma}

Before looking at the values of $I$ obtained with {\sc atoMEC} and the parametric potentials, it is important to have a rough idea of its possible values. A first approximation is given by a common model initially introduced by Bloch and used for instance by Deutsch \cite{Deutsch2016}: $I = 10.3~Z_T$ (eV). A slightly more complex approximation was published by Zimmerman \cite{Zimmerman1997}:
\begin{equation*}
    I (\mathrm{keV}) = Z_T \displaystyle\frac{0.024 - 0.013~ N/Z_T}{\sqrt{N/Z_T}}.
\end{equation*}
This will be our reference starting point for $I$. More than suggesting a formula for $I$, Zimmerman describes a full model for the stopping power with three formulas for the three components (see Eq. (\ref{threecom})). We use the formula for the ions (cores) contribution in order to systematically verify that this term is indeed negligible compared to the others. Zimmerman uses the Bethe formula with no correction terms for the bound electrons. \\
\begin{figure*}[ht!]
    \centering
        \centerline{\includegraphics[scale=0.27]{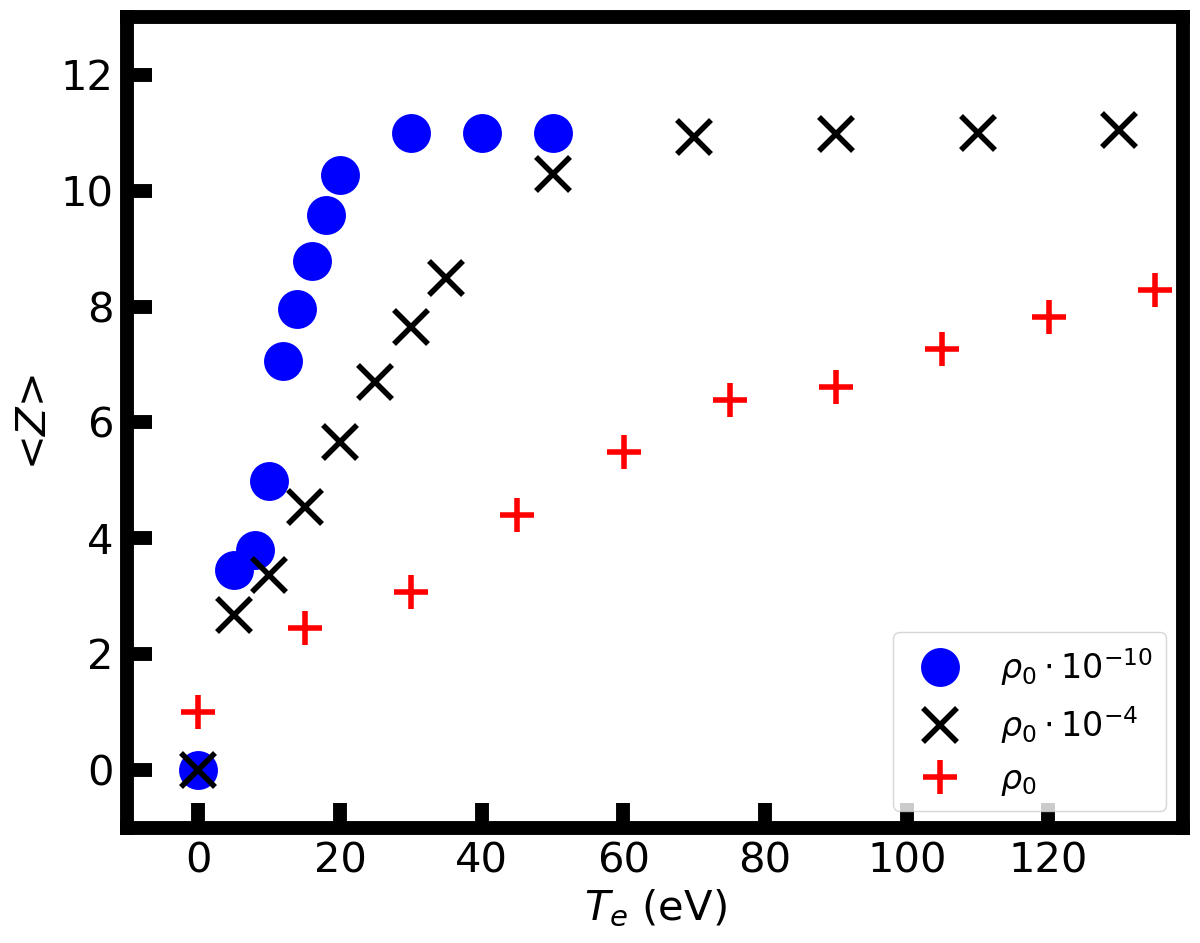} \includegraphics[scale=0.27]{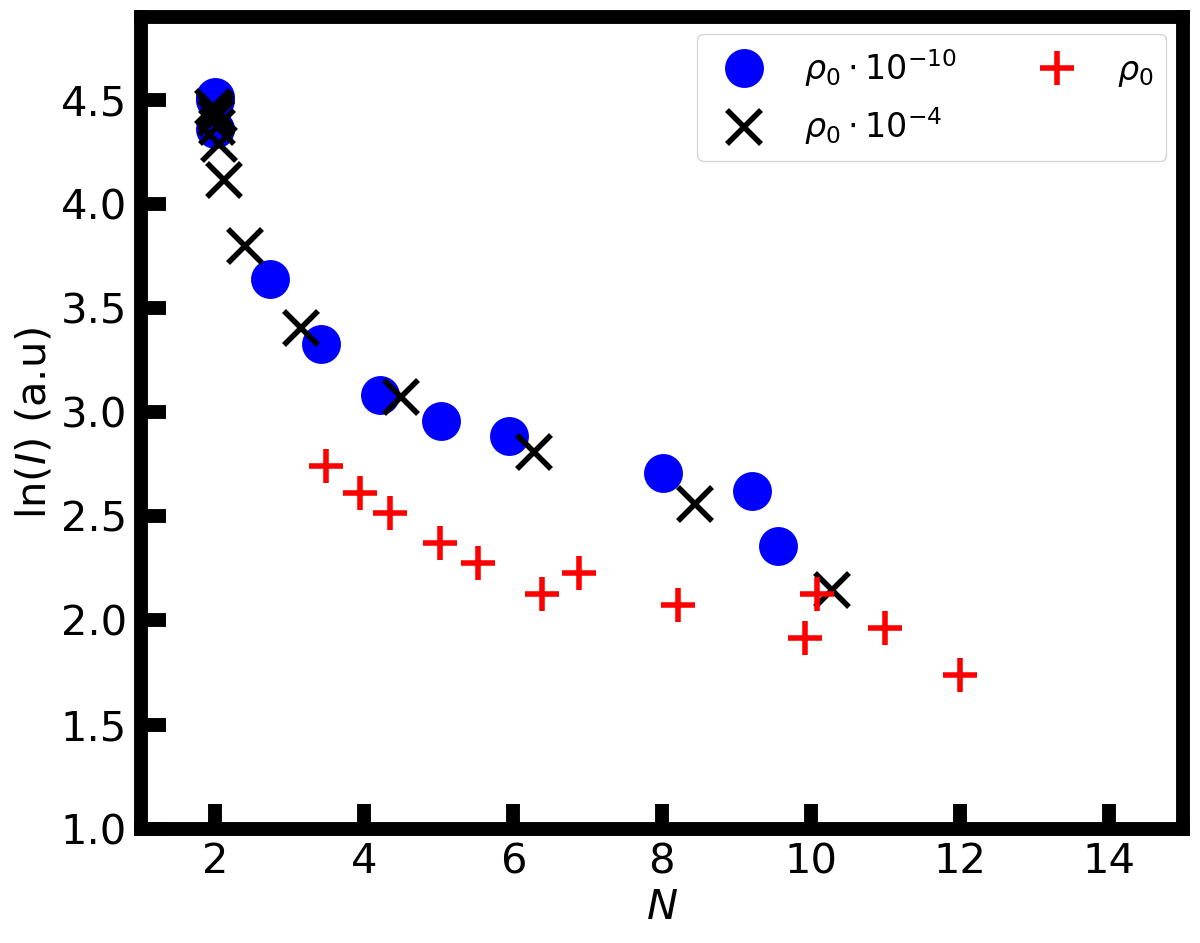}}
    \caption{On the left : Mean ionization from {\sc atoMEC} calculations for three densities and function of the electronic temperature $T_e$. On the right : Mean excitation energy of Al calculated using Garbet's method and {\sc atoMEC}'s data. For each data set, each point corresponds to a different temperature, hence a different number of bound electrons $N$.\label{I_atoMEC}}
\end{figure*}
In order to compare the mean excitation energies given by the parametric potentials and the average-atom code {\sc atoMEC}, we investigate the influence of different parameters of the code on $I$. We added a function to {\sc atoMEC} using the KS-wave-functions and following Garbet's method to find $K_0$ and $r_0^2$, thus to find $\rho(r)$ and $I$. First, {\sc atoMEC} calculations are performed for several densities and electronic temperatures ($T_e$) to determine the mean ionization $\langle Z\rangle$ for Al (Fig \ref{I_atoMEC}). The values of $I$ (Fig \ref{I_atoMEC}) are next calculated for different ionization degrees of Al using the converged SCF potential in {\sc atoMEC} calculations. For the same ionization degree, lowering the density tends to increase the value of $I$ until a maximum value, obtained around $10^{-4}$ g.cm$^{-3}$. In this study, the maximum value of $n$ is set to $n_{\mathrm{max}}=3$, due to the limited impact of high-$n$ subshells in the calculation of $r_0^2$.

\begin{figure*}[ht!]
    \centering
        \includegraphics[scale=0.3]{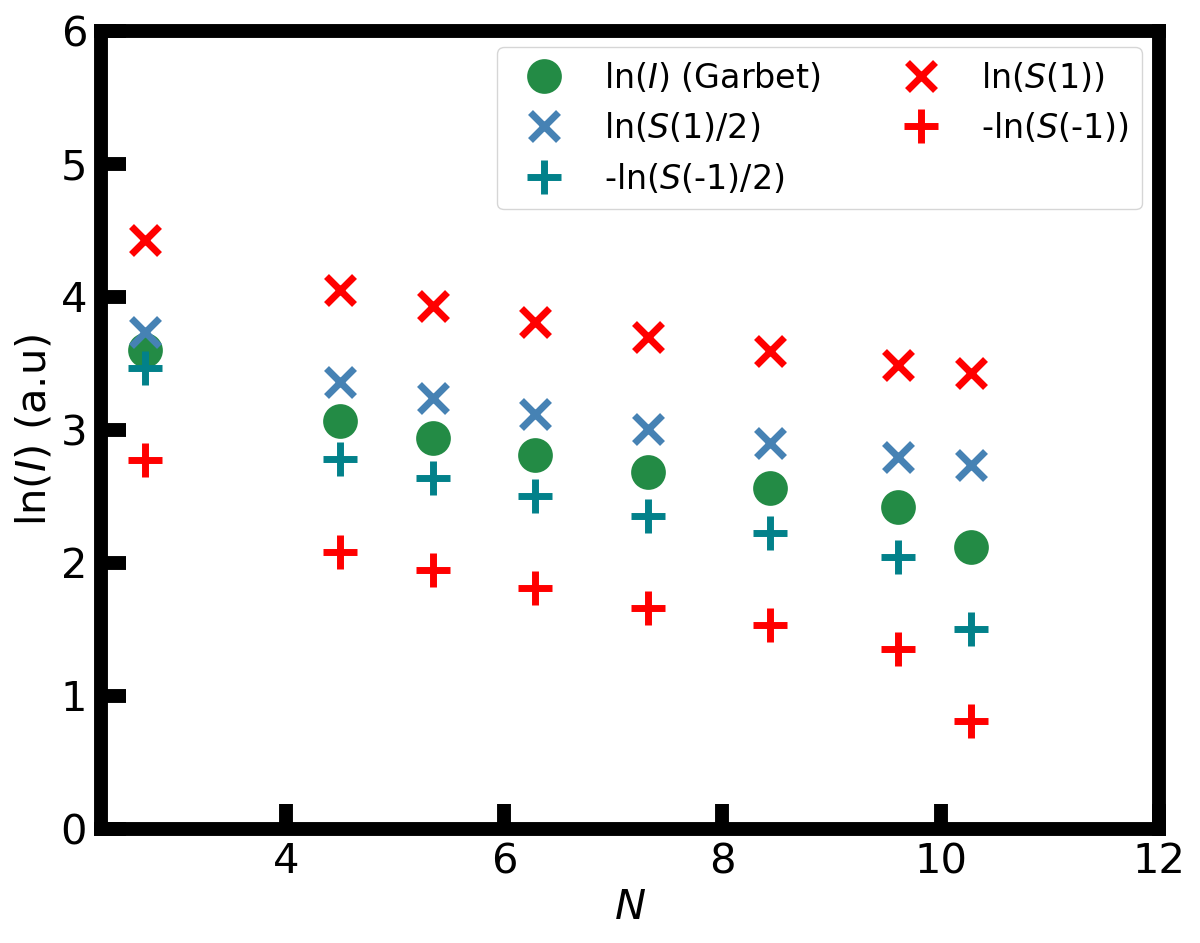}
    \caption{Mean excitation energy calculated using Garbet's method and {\sc atoMEC}'s data ($\rho_0 \cdot 10^{-4}$). Comparison with the bounds of the inequality $\ln \left[S(0)/S(-1)\right] \leq \ln(I) \leq \ln \left[S(1)/S(0)\right]$ taking $S(0)=1$ as Garbet suggests or $S(0)=2$, following the Thomas-Kuhn-Reiche rule. \label{InequalityCheck}}
\end{figure*}

In Fig \ref{InequalityCheck}, we can check the validity of the inequality (\ref{ln(I)}). $S(1)$ and $S(-1)$ are the mean kinetic energy and squared radius calculated with {\sc atoMEC}'s orbitals. Taking $S(0) = 1$, as suggested by Garbet, leads to a very weak inequality. Such a broad uncertainty cannot lead to errors below 8 \%. On the contrary, taking $S(0)=2$ as suggested in section \ref{I_Explained}, leads to a smaller gap, much more realistic with the error obtained. Nevertheless, if Garbet's method is approximate, it gives nevertheless a good idea of the behavior of $I$.\\

\begin{figure*}[t!]
    \centering
        \includegraphics[scale=0.25]{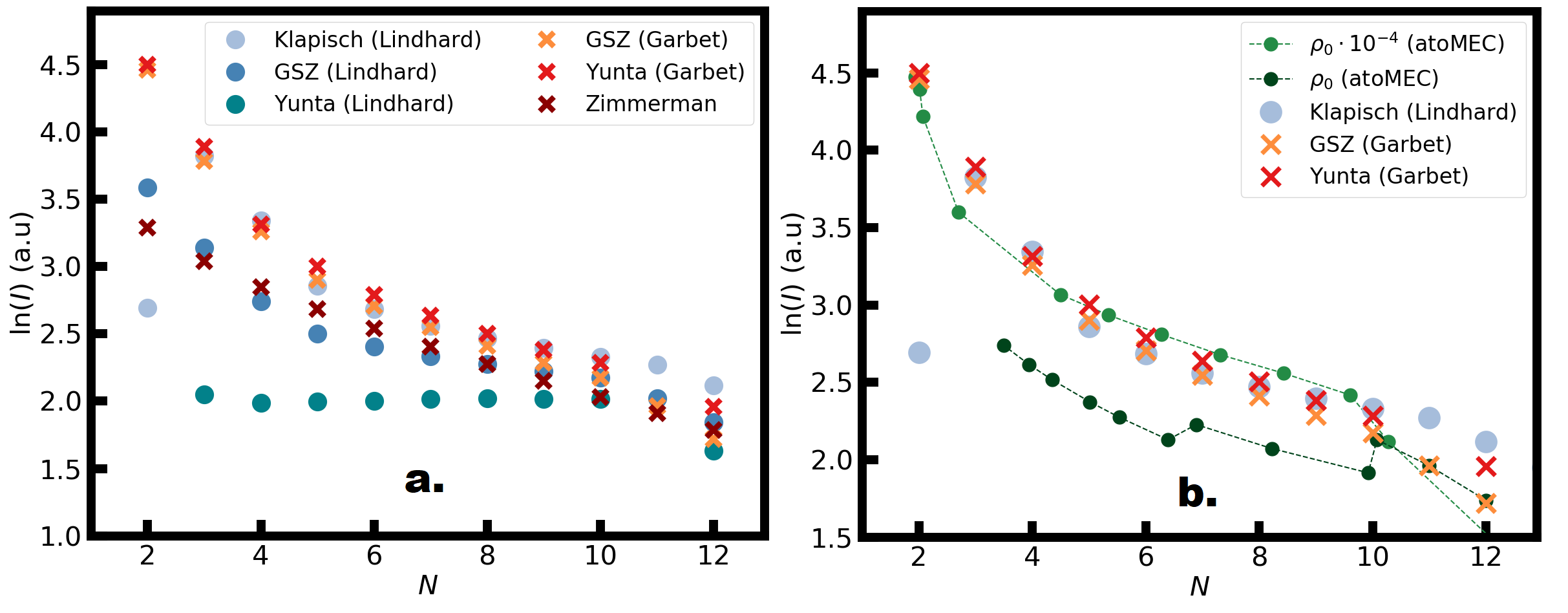}
    \caption{{\bf (a):} Comparison of mean excitation $I$ for Al obtained with Garbet's formula or the Lindhard formula for parametric and SCF potentials and the Zimmerman model. {\bf (b):} Comparison of the values of $I$ obtained with the Garbet formula and the potentials obtained from {\sc atoMEC} calculations and the parametric potentials. \label{Comparer_I}}
\end{figure*}

Now that both {\sc atoMEC} and the parametric potentials can be combined with Garbet's method. Their results are compared in Fig. \ref{Comparer_I}. It can be seen (Fig. \ref{Comparer_I} {\bf a.}) that $I$ follows the same behavior with the GSZ and Yunta potentials and also with the Zimmerman formula. As expected, $I$ decreases with $N$: considering that the energy of atomic shell $n$ is proportional to $1/n^2$ (quasi-hydrogenic scaling), the gap between two consecutive levels reduces when $n$ increases. As a result, the average excitation energy also decreases when $n$ increases, thus when $N$ increases. It can also be seen that GSZ and Yunta potentials lead to very similar results at low $N$ and that the differences between them logically increase with $N$. Note also that, as underlined before, the Martel parameters can lead to inconsistent values of $I$: the value for the Yunta potential at $N=11$ is not visible on the graph. We also plotted $I$ as defined by Lindhard and Scharff \cite{Lindhard1953}, who suggested setting
\begin{equation}
    \ln{I} = \frac{1}{Z_T} \int_{0}^{\infty} \ln\left[\gamma \hbar \omega_p(r)\right]\rho(r)4\pi r^2 \,\mathrm{d}r,
\end{equation}
where $\gamma = \sqrt{2}$ and $\omega_p(r)^2 = 4\pi e^2 \rho(r)/m_e$. This method is often considered as too crude in the literature (see Ref. \cite{Deutsch2016} for instance). It is however a very common model used for comparison. On the graph, it can be seen that both GSZ and Klapisch potentials lead to values consistent with the Garbet method (especially the Klapisch potential). On the contrary, certainly because its radial density does not vanish fast enough when $r$ tends to 0, the Yunta potential leads to unexpected results.

In Fig \ref{Comparer_I} {\bf b.}, the Garbet and Lindhard methods are compared to the values of $I$ obtained with Garbet's method and {\sc atoMEC}'s values. For ionized matter, the potentials lead to values of $I$ very close to the ones predicted by {\sc atoMEC} with a low density, the purpose being to make the {\sc atoMEC} results comparable to the ones from parametric potentials, defined for isolated ions. On the contrary, for neutral Al, {\sc atoMEC} with a solid-density seems closer to the values of the potentials. This could be explained by the specific treatment of valence electrons by {\sc atoMEC} which could display very low accuracy for neutral matter. 

\section{Total stopping power: calculation, results and comments}\label{sec4}

\subsection{Stopping power in cold matter}
 
The confidence in the reliability of the values of $I$ encourages us to compare our results obtained with the Bethe formula to two reference codes for cold matter. The first one, developed by the National Institute of Standards and Technology (NIST), is called {\sc PSTAR} or {\sc ASTAR} (for protons or $\alpha$-particles used as projectiles) \cite{PSTAR}. At high projectile energies, it involves the Bethe formula with correction terms. However, at low energies (under around $\approx$ 0.5 MeV for protons), it uses fits over experimental data. The expected error at high energies is claimed to be lower than 2 \%, but at low energies, it is said to be up to 10 \% at 100 keV and up to 30 \% at 1 keV. Even if these errors sound significant, they only impact low-energy projectiles: limited amounts of energy can be transferred by such projectiles.

The second reference is the {\sc SRIM} code, developed by Ziegler {\it et al.} \cite{SRIM}. It is based on a Monte-Carlo calculation in the binary-collision approximation: the projectile experiences independent binary collisions with atoms while passing through materials. It follows a straight path between two events and both the distance before the next collision and its impact parameter are randomly chosen in a weighted distribution depending on the density. This method is expected to be accurate for projectiles with energies between 10 eV and 2 GeV. 
 \\

\begin{figure*}[t!]
    \centering
        \includegraphics[scale=0.27]{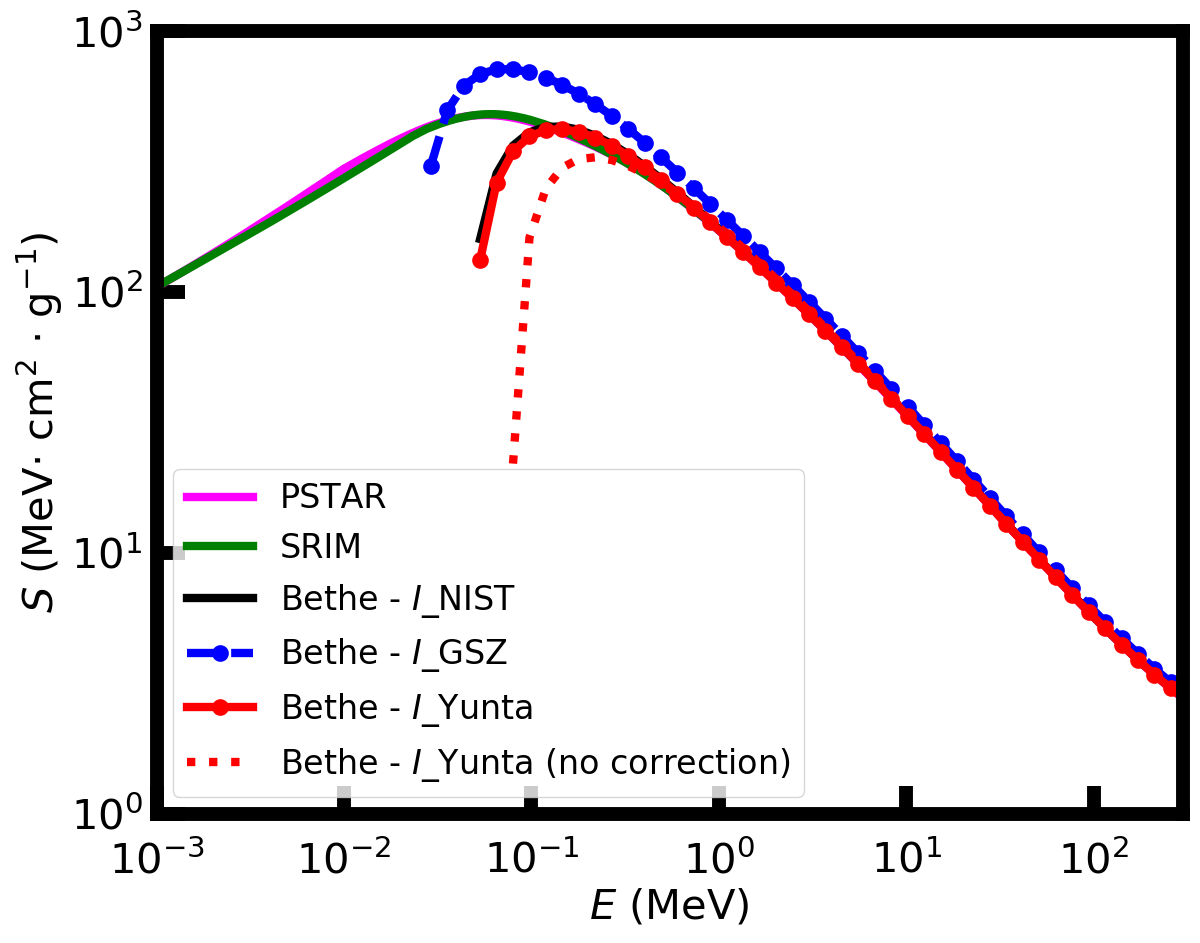}
    \caption{For cold matter of neutral Al , the stopping powers of the {\sc PSTAR} and {\sc SRIM} codes and Bethe formula are compared to different versions of the Bethe formula (with $I$ as defined by the GSZ potential, by the Yunta potential and taken from the NIST, respectively) \label{PSTAR}}
\end{figure*}

In Fig. \ref{PSTAR}, we represented the total stopping power in (cold) neutral aluminum as a function of the incident proton kinetic energy. A first remark is that the {\sc PSTAR} and {\sc SRIM} codes lead to very similar results, supporting the idea that they can be used as references. As expected, the Bethe formula fits the references over a limited range of energies. First, at low energies, the Bethe formula cannot describe the physics of the stopping power: as expected, the {\sc PSTAR} starts using fits over experimental data under 0.5 MeV and the Bethe formula is not accurate anymore. Nevertheless, the addition of corrections to the Bethe formula leads to a significant improvement and adds half a decade to the range of acceptable values. Then, at high energies, above 100 MeV, the density-effect correction starts to become significant. We believe the absence of this term in our calculations is responsible for the difference with the {\sc PSTAR} code. Overall, the Bethe formula with correction terms covers a range of energies between 0.3 and 100 MeV, which is enough for us. Significant differences can be observed between the Yunta and the GSZ potentials. The addition of the Klapisch potential would not be of any use, as we fixed its parameters to reproduce the GSZ values of $K_0$ and $r_0^2$. For neutral Al, the Yunta potential gives a mean excitation energy much closer to the one used by the NIST (166 eV) than the GSZ potential. The same observation can be done with Si instead of Al. Even though this result sounds counter-intuitive as the Martel parameters for the Yunta potential lead to some inconsistencies, such as an unexpected shape of $\rho(r)$, the Yunta potential seems to lead to better values of $I$ than the GSZ potential for neutral matter (note that even if the NIST can be held as a reference, getting close to its values for $I$ does not necessarily means getting close to its predicted stopping power).

\subsection{The Maynard-Deutsch-Zimmerman formula for highly ionized matter \label{Section_Maynard}}

In fully ionized matter, the stopping power is due to the gas of free electrons. As a consequence, the Bethe formula cannot be used anymore and we resort to the Maynard and Deutsch approach \cite{Maynard1985}, considering a homogeneous dense electron fluid and use the Random Phase Approximation (RPA) to calculate the dielectric function. The RPA is valid for a range of $T_e$ and $\rho$, such as $\chi^2/(1+T)\ll 1$ with $\chi^2 = 1/(\pi q_F a_0)$. $q_F$ being the Fermi wave number. This inequality is verified in most systems involving nuclear fusion, from high-temperature plasma of tokamaks to dense plasma in ICF systems. Considering only elastic processes for the projectile-electrons interactions, it is possible to use the framework of linear response theory, $\textbf{J}_{\mathrm{ind}} = -i (\omega /4\pi) (\epsilon(\textbf{q},\omega)-1)\,\textbf{E}(\textbf{q},\omega)$ with $\epsilon$ the dielectric function, $\textbf{J}_{\mathrm{ind}}$ the induced current vector, and ${\bf E}$ the electric field. The calculation of $\epsilon$ is explained in Ref. \cite{Gouedard1978}. In the RPA, electrons are only influenced by the total electric potential (the sum of an external potential and a screening potential). Considering a Fourier-transformed many-electron Hamiltonian, the RPA assumes that the contribution to $\epsilon({\bf k})$ of random momentum transfers ${\bf k+q}$ averages out. Consequently, only the total potential for the wave number ${\bf k}$ matters. The resulting $\epsilon$ is sometimes called the Lindhard dielectric function. Adding the Born approximation (in scattering processes, one only considers waves resulting of none or one interaction), the stopping power can be expressed as
\begin{equation*}
    S = \displaystyle\frac{2}{\pi} \left(\frac{Z_{\mathrm{eff}}\, e}{V_p}\right)^2 \int_0^{\infty} \displaystyle\frac{\,dq}{q}  \int_0^{\infty} \omega~\text{Im}\left[\displaystyle\frac{1}{\epsilon(q,w)}\right]\,\mathrm{d}\omega.
\end{equation*}
Maynard and Deutsch exhibit simple limits for the stopping number. At high temperature ($T/T_F \to \infty$), if $x=V_p/V_{th}$ with $V_{th}\approx V_F a_0$ ($V_F$ is the Fermi velocity), they show that for $x\gg 1$ (high velocities):
\begin{equation*}
    L_e = \ln\left(\frac{2m_eV_p^2}{\hbar \omega_p} - \frac{3}{2x^2} - \frac{15}{8x^4}\right)
\end{equation*}
and for $x\ll 1$:
\begin{equation*}
    L_e = \frac{2\,x^3}{3 \sqrt{\pi}}~\ln \left(\frac{1}{4 \delta e \gamma}\right)
\end{equation*}
with $\gamma = \exp (0.517)$ and $\delta^{1/2} = \hbar/\left(4m_e V_{th} \lambda_D\right)$. The authors also suggest a bridge function to link the two limits. They define $y$ as $V_p/V_e$ with $V_e$ the most probable electron speed and propose
\begin{equation*}
    L_e = \ln\left(\frac{1}{\sqrt{2\delta}}\right) \left[\text{erf}(y) - 2\frac{y}{\sqrt{\pi}}\,e^{-y^2}\right]\,P(y),
\end{equation*}
where $P(y)$ is a fraction of polynomials of $y$. However, it seems to us that $P(y)$ is not correctly defined to recover the proper limits and we therefore prefer to use the Zimmerman function, which relies on the work of Maynard and Deutsch, only replacing the coefficients and degrees of the polynomials of $P(y)$.

It should also be stated that the limit $T/T_F \to \infty$ does not restrict the scope of the present work, first because such low temperatures are not considered, second because at low temperatures, the ionization of matter is small, which is also the case of the free electrons' stopping power is also limited. \\

\begin{figure*}[t!]
    \centering
        \includegraphics[scale=0.33]{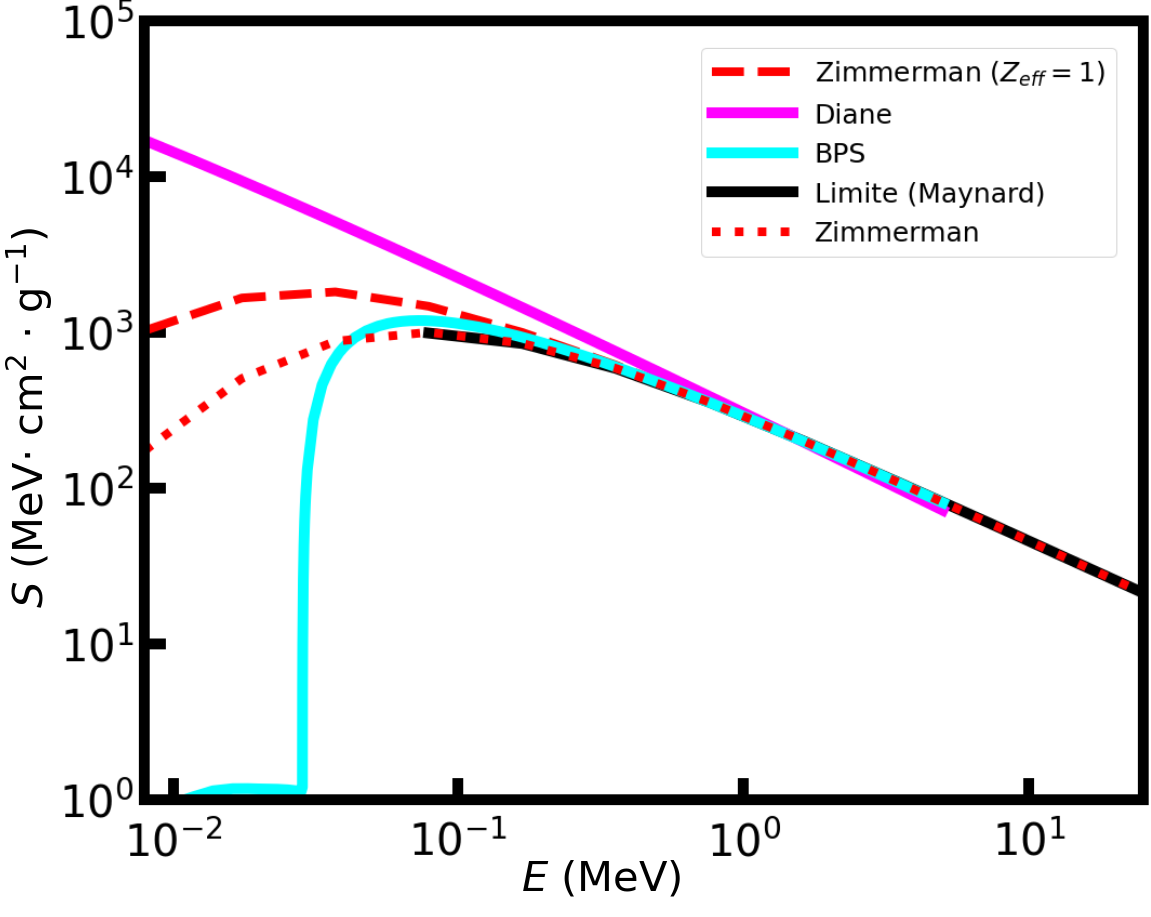}
    \caption{In fully ionized Al at 1000 K and for an incoming proton, comparison of the stopping power of the free electrons as described by the reference codes, the Maynard-Deutsch-Zimmerman limit function with and without ($Z_{\mathrm{eff}}=1$) the Brown correction of the projectile charge. \label{Maynard}}
\end{figure*}

For ionized matter, we use two main reference codes. The first one relies on the Brown, Preston and Singleton ({\sc BPS}) approach \cite{BPS}. It is based on an adaptation of the dimensional continuation method of gauge-invariant quantum field theory to dense plasma. For a non-relativistic particle, the long-distance excitations of the plasma are described by the Lenard-Balescu equation and the short-distance collisions of the plasma particles are modeled by the Boltzmann equation for Coulomb scattering. Such a modeling works for all charges, velocities or masses of the projectile. However, it is assumed that the plasma is not strongly coupled: the coupling parameter $g = e^2K_D / 4\pi T$ with $K_D$ the Debye wave number, must be small. Such an approach is not fully appropriate for lower beam velocities as shown experimentally by Cayzac {\it et al.} \cite{Cayzac2017}. Moreover, it is worth mentioning that the multi-dimensional analysis of BPS to include hard collisions is not necessary as shown by Gericke and Schlanges \cite{Gericke1999}, generating data that compare well with Cayzac {\it et al.}'s experimental data and simulations.

The second one is called {\sc Diane} \cite{Diane}. {\sc Diane} is a code developed at CEA to model the stopping power of a fully ionized plasma. It has been used in particular in the context of the slowing down of high-energy charged particles ($T>1$ MeV for protons) in fully ionized light element plasmas. As {\sc BPS}, it is based on the Lenard-Balescu model, which considers binary collisions (Coulomb scattering) for short-range interactions and models long-range interactions using screened potentials. However, {\sc Diane} involves limit formulas for high velocity projectiles, which makes the calculation faster. \\

In Fig \ref{Maynard}, all different models are compared for an electron gas in conditions such that the Maynard and Deutsch formula is valid and the coupling parameter $g$ is such that $g\ll 1$. Thus, {\sc BPS} should give consistent results. On the graph, only the high-velocity asymptotic forms plotted, as the low velocities do not appear in the range of considered energies. A first remarkable aspect is that all models lead to the Maynard limit at high energies. Consequently, the Zimmerman function seems to be a suitable model for energies above 0.5 MeV. However, similarly as the case of cold matter, significant differences are visible for lower energies. The {\sc BPS} and {\sc Diane} codes display opposite behaviors. This is due to the limit formula used by {\sc Diane} for high-velocity projectiles, which is not expected to be valid at low velocities ($E<$ 0.1 MeV). \\

This example also illustrates the impact of the Brown formula (see Eq. (\ref{bro})) for $Z_{\mathrm{eff}}$, the effective charge of a proton. As previously stated, we take $Z_p=1$ for protons. However, if the Brown formula is used, its influence is visible at low velocities: it changes the final value of the stopping power by a significant amount. In the frame of this work, we consider energies above 0.3 MeV. Thus, taking $Z_p=1$ instead of $Z_\mathrm{eff}$ does not lead to a significant difference. Nevertheless, if the {\sc BPS} code is considered as accurate, then the actual charge must be between $Z_p=1$ and the Brown formula, according to Fig. \ref{Maynard}. Considering the Bethe formula and cold matter, the same observation can be made: the difference in stopping power between $Z=1$ and $Z_\mathrm{eff}$ is limited for protons.

\subsection{Total stopping power in partially ionized plasma}

\begin{figure*}[ht!]
    \centering
        \includegraphics[scale=0.27]{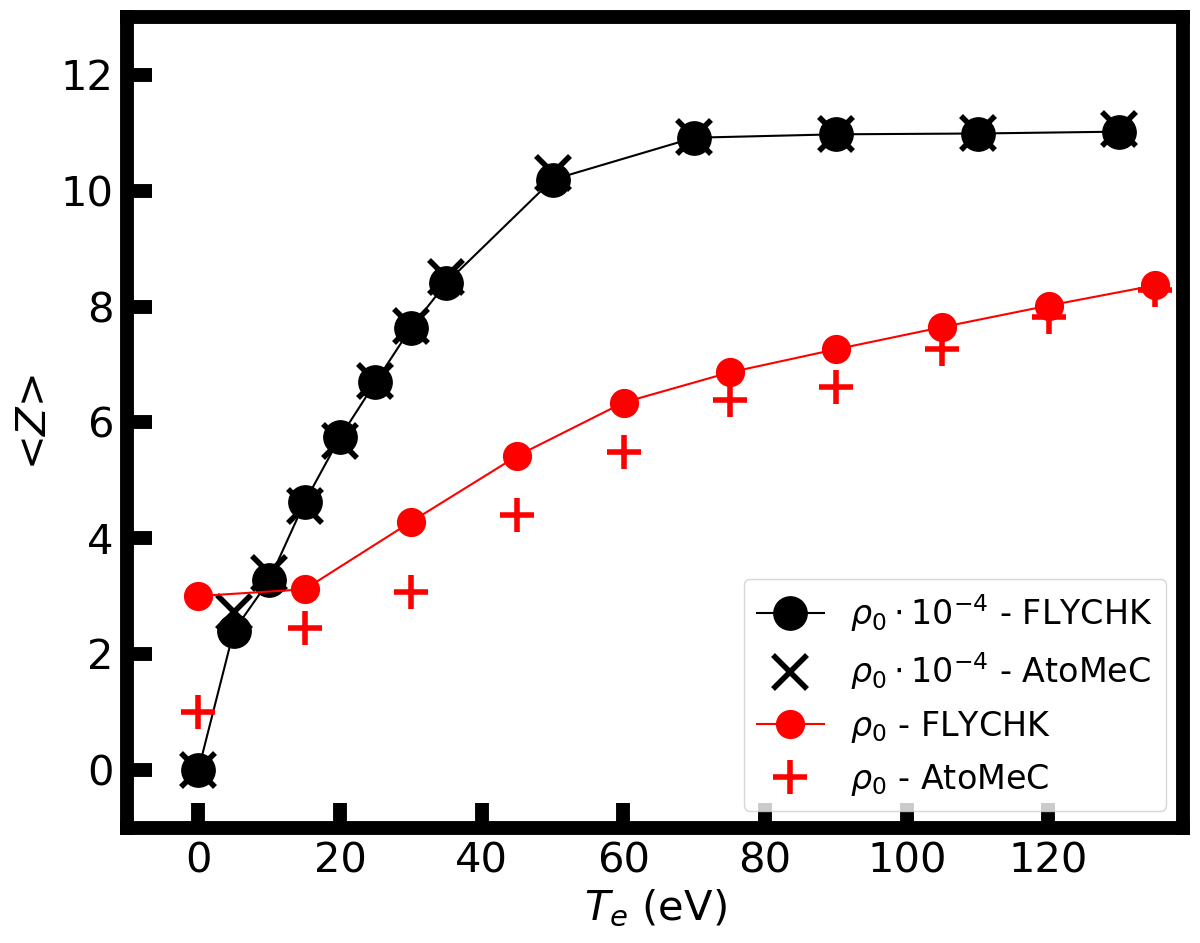}
    \caption{ {\bf (a).} Mean effective charge of Al as a function of the temperature, using {\sc FLYCHK}  or {\sc atoMEC} with $\rho = \rho_0$ or $\rho = \rho_0 \cdot 10^{-4}$, $\rho_0$ being the standard solid-state density.} \label{FLYCHK}
\end{figure*}

At this stage, we have a reliable model for bound electrons in cold matter using the Bethe formula and the parametric potentials, as well as a good description of the stopping power of a gas of electrons, using the Zimmerman formula. In the latter case, we also checked that the contribution of the ions (cores) to the stopping power is negligible over the range of energies. In order to describe a real plasma, we now need to know the degree of ionization of the matter as a function of the density and the temperature. This relation can be found using the Thomas-Fermi model. As shown in Ref. \cite{Drake2018} (pp. 78-79), the mean charge $Z$ of an ion with atomic number $Z_T$ can be expressed as $Z = f(x)~Z_T$ where $x$ depends both on $T_e$ and $\rho$. However, the potentials are defined only if $Z$ is an integer. Thus, we use the {\sc FLYCHK} code \cite{FLYCHK} available on the NIST website to get the ionization distribution (i.e., the ionic fractions) of the atoms. {\sc FLYCHK} uses schematic atomic structures and, by considering collisional and radiative processes, it solves the rate equations. In Fig \ref{FLYCHK}, the mean ionization calculated by {\sc atoMEC} and {\sc FLYCHK} in order to make sure that they lead to close results. With solid-density, both codes are similar for all temperatures. On the contrary, at low densities, {\sc FLYCHK} tends to describe Al as ionized (Al$^{3+}$) at room temperature, which is less the case with {\sc atoMEC}. As we previously highlighted that the number of shells considered with {\sc atoMEC} impacts the calculated $r_0^2$, we calculated the mean ionization using $n_{\mathrm{max}}=3$ or $5$. The results are very close, considering 3 or 5 orbitals leads to the same ionization with {\sc atoMEC}, which supports the idea that the variations of $I$ are dominated by the distribution of electrons over the most loosely-bound shells. 

\begin{figure*}[ht!]
    \centering
        \centerline{\includegraphics[scale=0.27]{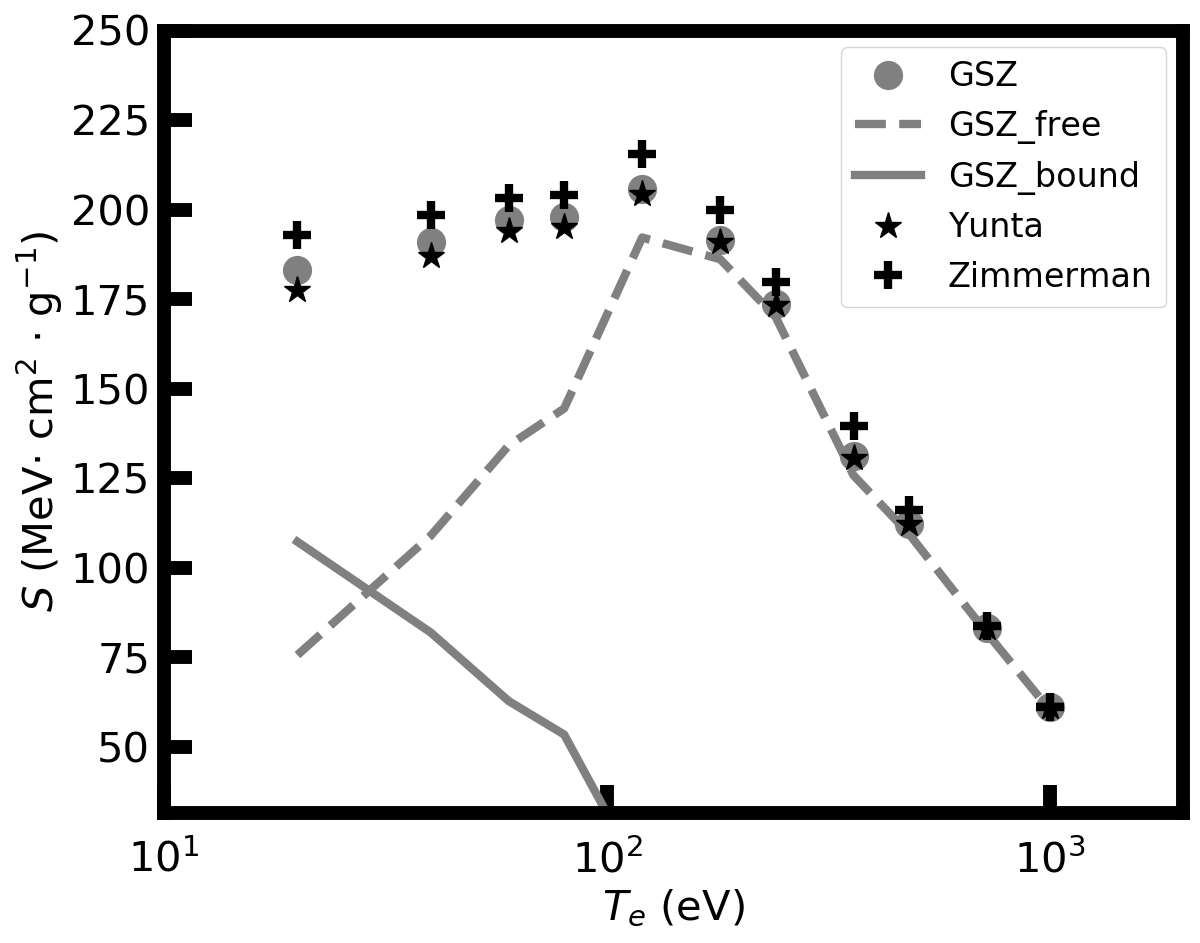} \includegraphics[scale=0.27]{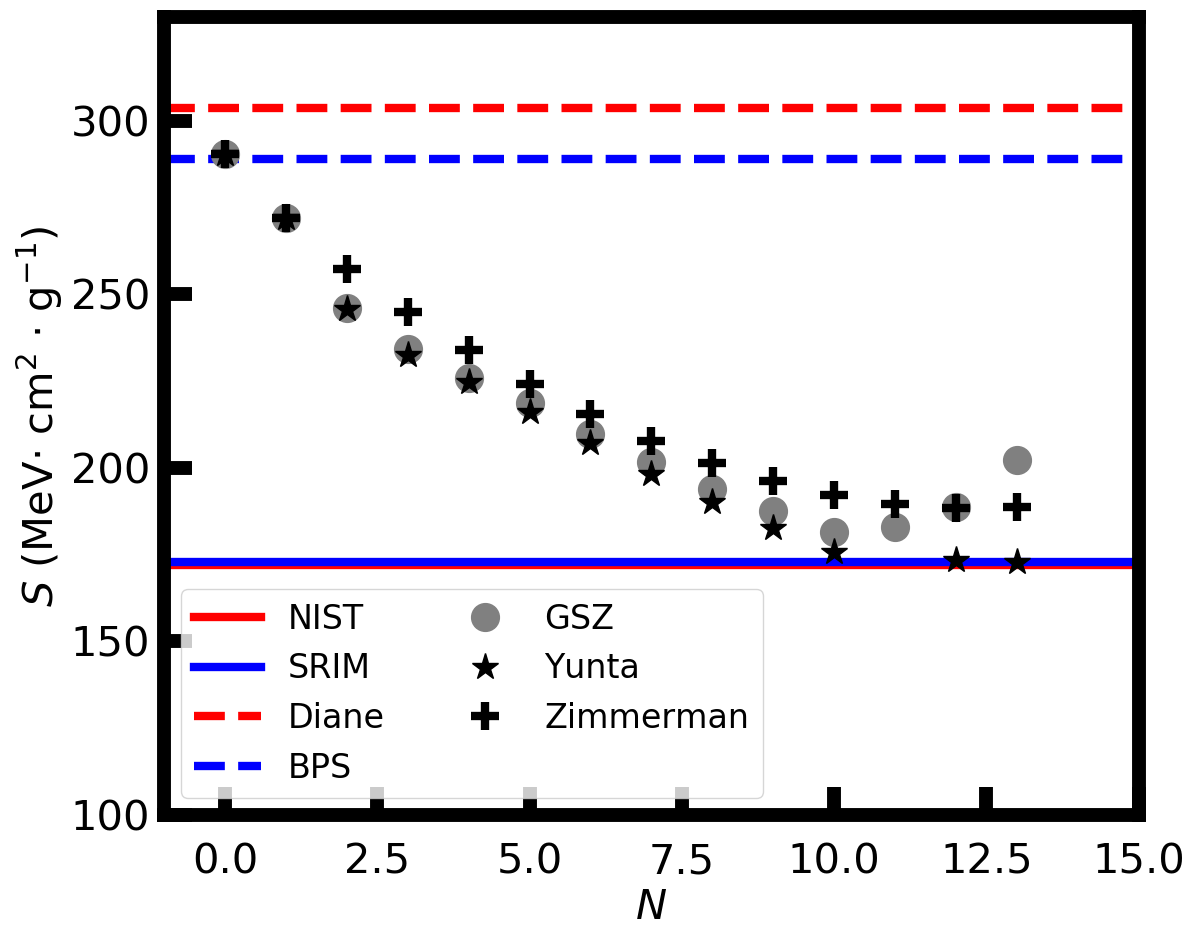}}
    \caption{{\bf On the left:} Total stopping power of a proton with $E$=1 MeV in Al at solid density, as a function of the temperature. The distribution of the ionization is calculated with {\sc FLYCHK} and the contribution of the bound electrons is calculated with the Bethe formula with the GSZ potential ($\bullet$) or the Yunta potential ($\star$) or with the Zimmerman function ($+$). The contribution of the free electrons (dashed line) is calculated with Zimmerman's function. For the Yunta potential, the contribution of the bound electrons is also plotted separately (continuous line). {\bf On the right:} For the same system, total stopping power as a function of the ionization and for $T_e= 0.1$ eV. The same models are used (GSZ ($\bullet$), Yunta ($\star$) or Zimmerman($+$)). Limits for neutral Al are plotted with continuous lines: the {\sc SRIM} (red) and {\sc PSTAR} (blue) codes. For fully ionized matter, the {\sc Diane} (red) and {\sc BPS} (blue) codes are represented with dashed lines. \label{S_Plasma}}
\end{figure*}

By definition, as $\ln(I)$ has a more physical meaning that $I$, let us consider, for a distribution of ionization stages with probabilities $p_n$ for each ionization level $I_n$, let us consider $\ln(I) = \sum_n p_n \ln(I_n)$. For a proton in Al, the result can be seen in Fig \ref{S_Plasma} {\bf a.}. The stopping power reaches a maximum for a temperature of around 120 eV (most probable ionization: Al$^{10+}$). This maximum corresponds to the maximum stopping power of free electrons, as the contribution of bound electrons is rather limited at this temperature. The contribution of bound electrons is significant at low temperatures, for which the mean ionization is under 10. \\

In order to better describe the influence of the electrostatic potentials on the stopping power $S$, Fig \ref{S_Plasma} {\bf b.} shows $S$ as a function of $N$ at fixed $\rho$ and $T_e$. This representation emphasizes on the fact that the models investigated here bridge the previously explored limits of neutral and fully ionized matter. The Yunta potential best fits the reference codes for neutral matter and the Zimmerman function for free electrons correctly reaches the fully ionized matter asymptotes. Between these limits, the potentials display slightly different behaviors. However, as pointed out by Gauthier \textit{et al.} \cite{Gauthier2013}, no experimental data actually exist in dense plasma at temperature higher than 1 eV. Since then, very few experiments have been led, due to the fact that very few facilities can produce plasma and high energy protons. As a result, it is almost impossible to compare our models to any experimental data and to select a potential among them.

Isochoric heating experiments using proton beams on solid targets were carried out on laser facilities \cite{Mancic2010,Hoarty2012}. In very recent experiments conducted at LULI (Laboratoire pour l’Utilisation des Lasers Intenses, Palaiseau, France) \cite{Rassou2025}, protons generated via target normal sheath acceleration (TNSA) from a gold foil irradiated by a short-pulse laser were used to deposit energy into aluminum or copper foils initially at room temperature and solid density. This energy deposition drives the material into the warm dense matter regime, with rear-surface temperatures ranging from 1 to 5 eV. Temperature measurements were obtained using streaked optical pyrometry, while the proton beam characteristics were determined using a Thomson parabola spectrometer. To reconstruct the initial proton distribution prior to interaction with the target, high-energy proton generation by TNSA was modeled. In radiation-hydrodynamic simulations, proton heating was treated using a Monte Carlo transport module for charged particles, and accurate energy deposition calculations required the use of stopping power models. These ongoing experimental efforts are expected to provide critical insights into the validity and limitations of the theoretical modeling presented in this work.

\section{Conclusion}

As a whole, we presented a formalism aiming at reproducing the stopping power functions for neutral and fully ionized matter of the reference codes {\sc Diane}, {\sc BPS}, {\sc SRIM} and {\sc PSTAR}. Moreover, we suggested a function to bridge the neutral and ionized matter limits and calculate the stopping power of a partially-ionized plasma, using both the Bethe formula and the Maynard or Zimmerman approach for the free electrons. In particular, we investigated a fully analytical approach of the problem, using and comparing different forms for the electrostatic potential (parametric or SCF average-atom). We highlighted the interest of the Klapisch potential and the necessity of defining a full set of parameters using an energy-minimization code. We also presented the Martel parameters for the Yunta potential as leading to the best approximation of the mean excitation energy, even if the Yunta potential leads to an unexpected behavior for the radial density. However, the lack of experimental data represents a strong limit to this work as we cannot clearly state that a potential is better than the others. Still, we underlined the proximity between the parametric potentials and the average-atom potential of the {\sc atoMEC} code. Using {\sc atoMEC}, we pointed out the limitations of average-atom codes, distributing electrons over all available shells, including the most loosely-bound ones. Thus, {\sc atoMEC} and Garbet's approaches are rather different. This work was also an opportunity to test Garbet's method to reach the mean excitation energy, $I$, validate the use of the Virial theorem and the consistency of the approximation of $\ln (I)$ using sum rules. Although several approximations need to be questioned (the form of the effective charge $Z_{\mathrm{eff}}$ for light ions is not perfectly defined, the corrections for the Bethe formula are limited in range, \textit{etc.}) and although the resolution of the Schr\"odinger equation for the parametric potentials remains an issue in the context of an (as much as possible) analytical approach, we believe that the formalism discussed here should be helpful and yield realistic values of the stopping power. It would be rewarding to extend the present analysis to arbitrary coupled plasmas by including plasma density effects (such as ionization potential depression, see for instance Refs. \cite{Ecker1963,Stewart1966,Preston2013,Crowley2014,Pain2022}), allowing mixing between bound- and free-electron stopping powers. In the present work we focused on the plasma target (its mean ionization degree and electronic structure), but the issue of the ionization of the projectile is important as well. Accounting for the latter requires a different approach than the one we used for the plasma target, because for the projectile one needs to account for the temporal evolution of the mean ionization with time, which implies to model the atomic kinetics. Although we consider light projectiles (hydrogen, helium) which are mostly fully stripped, this is a topic we intend to investigate in a future work. It can be important, for instance, in the framework of heavy-ion fusion. Finally, experiments in the range of plasma temperature and density corresponding to partial ionization are lacking, and we hope that the laser M\'egajoule, in conjunction using the Petawatt beam, will provide high-quality experimental data.


\section*{Author declarations}

\subsection*{Conflict of interest}

The authors have no conflicts to disclose.

\subsection*{Data availability}

The data that support the findings of this study are available from the corresponding author upon reasonable request.

\appendix

\section{Corrections to the Bethe formula}\label{appA}

First, the shell correction term ($C/Z_T$) tackles the issue of low projectile velocity. The Bethe theory relies on the assumption that we have $V_p>V_e$, with $V_e$ the velocity of the bounds electrons. As this is not always the case, the shell correction term adapts the formula by considering different velocities associated, in a classical interpretation, to the bound-electron subshells. Then, the density effect term ($\delta/2$) takes into account the polarization effects in the target, which reduce the stopping power since the ion’s electromagnetic fields are reduced by the dielectric constant of the target medium \cite{Ziegler1999}. Finally, $L_0$ becomes :
\begin{equation*}
    L_0 = \ln\left[\frac{2m_e \beta^2 c^2}{I (1-\beta^2)}\right] - \beta^2 - \frac{C}{Z_T} - \frac{\delta}{2}
\end{equation*}
In the present work, we focus on projectiles with energies below 100 MeV, and neglect the density-effect term (see the values for Al calculated by Bichsel \cite{Bichsel2002}). For the shell correction, we resort to the rather simple Fano \cite{Fano1963} expression: 
\begin{equation*}
    \frac{1}{2}\frac{C}{Z_T} = \frac{Z_T^{0.4}}{2x} + \frac{bZ_T}{x^2} \qquad \text{ with } \qquad x=\frac{V_p^2}{v_0^2 Z_T},
\end{equation*}
with $b$ between 4 and 5, slowly increasing with $Z$. $v_0$ represents the Bohr electron velocity.

In 1956, Barkas discovered a difference in the range (the distance traveled in a material) of positive and negative particles. This is due to the attraction of the electrons by positive particles toward their trajectories \cite{Barkas1956,Barkas1963,Maynard1982,Andersen1989,Moller1997}. The electrons leave their equilibrium positions and scattering events become more common and more intense. The opposite phenomenon happens with the repulsion by negative particles. 
Bichsel notes that no complete theory of $L_1$ is available and that empirical expressions should be used. Ziegler proposed the form \cite{Ziegler1999}:
\begin{equation*}
    L_1 = \frac{L_\mathrm{low} L_\mathrm{high}}{L_\mathrm{low} + L_\mathrm{high}},
\end{equation*}
where $L_\mathrm{low} = 0.001\,E$ and $L_\mathrm{high} = (1.5/E^{0.4}) + 45,000/Z_T \cdot E^{1.6}$. $E$, in keV/u (u denoting the atomic mass unit), is the kinetic energy of the projectile. Ziegler notes that the expression is not reliable below 1 MeV/u \cite{Ziegler1999}. Bichsel points out that the use of his form for energies below 0.8 MeV/u and above 5.6 MeV is an extrapolation which may give unreliable results. Nevertheless, both models show good consistency above 0.3 MeV/u, which is enough for the range of energies we investigate.\\

Bloch compared the quantum mechanical modeling of Bethe to Bohr's classical approach of the stopping power of Bohr \cite{Ziegler1999}. He found a rather good agreement for small impact parameters between the two formulalisms. He showed that higher-order terms were necessary for larger impact parameters, in particular the $Z^4$ term. This term is commonly expressed as \cite{Protontherapy}:
\begin{equation*}
    L_2 = -y^2 \sum_{l=1}^{\infty} \frac{1}{l(l^2+y^2)} \qquad \mathrm{with} \qquad y=\frac{Z_T}{137\beta}.
\end{equation*}
Calculating the sum by including successively $10^2$ and $10^4$ terms, the difference is of less than $0.1\%$. As a result, $10^4$ terms seems to be a good enough approximation of the Bloch correction. 

\section{Determination of the parameters of the Klapisch potential}\label{appB}

\begin{figure*}[ht!]
    \centering
        \includegraphics[scale=0.34]{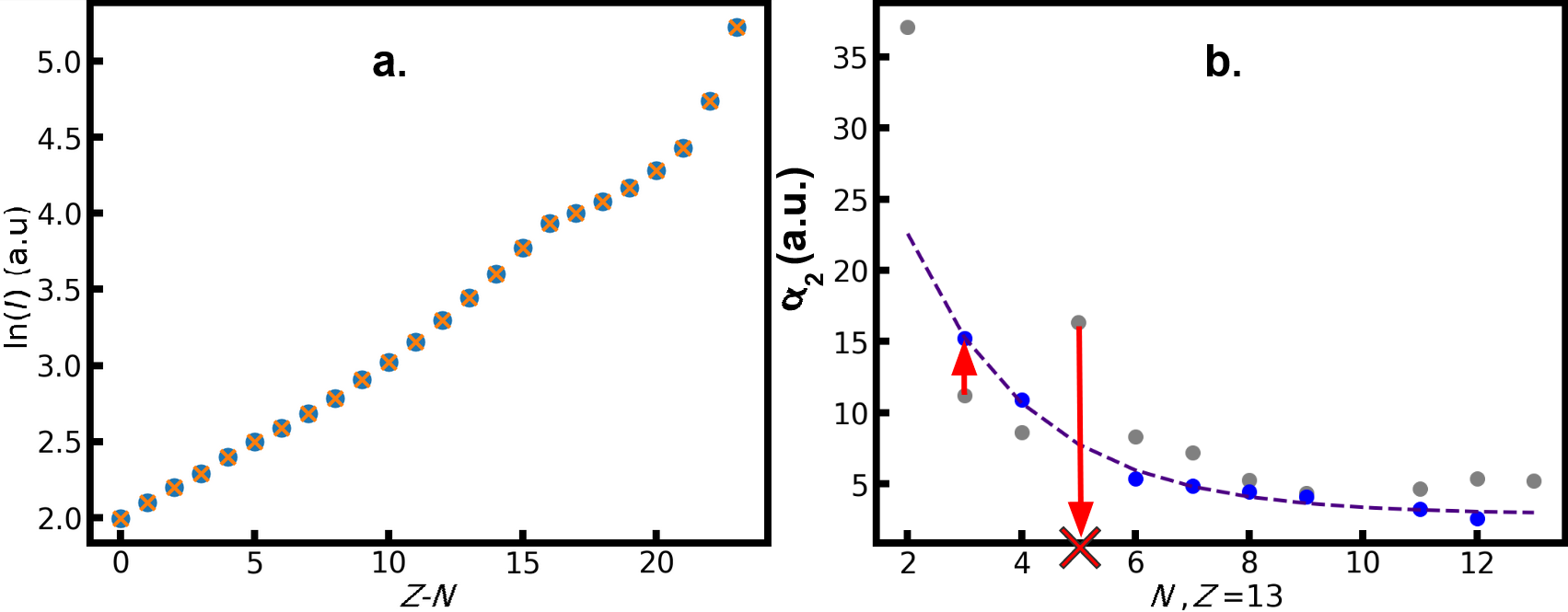}
    \caption{{\bf (a):} Comparison of $I$ as obtained using the GSZ potential ($\bullet$) and the Klapisch potential ($\times$) using the parameters defined after the first minimization over the values of $I$. 
    {\bf (b):} Parameter $\alpha_2$ of the Klapisch potential. Using the parameters defined by the first minimization (grey) as an initial estimate, we minimize the equations over $K_0$, $r_0$ and $r^2_0$. If the resulting parameter $\alpha_2$ is greater than zero, it replaces the previous one (blue). If not, it is rejected. An exponential fit over the newly defined parameters is plotted (dashed line). \label{First_Step}}
\end{figure*}

An interesting idea to get the parameters of Klapisch's potential is to compare $I_{\mathrm{GSZ}}$ and $I_{\mathrm{Klapisch}}$ as this is the quantity of interest for stopping powers. However, setting the three parameters for a $(N,Z_T)$ couple at once solely with the equation $I_{\mathrm{GSZ}}(N,Z_T) = I_{\mathrm{Klapisch}}(N,Z_T)$ seems rough. Hence, just as Garvey considered the GSZ parameters as linear functions of $Z_T$, We write the parameters as $\alpha_{i,N}(Z_T) = \alpha_{A,N} + Z_T\cdot \alpha_{B,N}$. In this way, we use a least squares approach (\verb!scipy.optimize.leastsq!) to find the suitable parameters $\alpha_{A}$ and $\alpha_{B}$ for each $N$. After a first minimization over a limited range of $Z_T$, we take the obtained values as an initial estimate for a second minimization over all possible $Z_T$. The results are showed in Fig \ref{First_Step} {\bf a.} $I_{\mathrm{GSZ}}$ and $I_{\mathrm{Klapisch}}$ are very well superimposed but when looking closely at the parameters of Klapisch's potential, it appears that they do not follow all physical conditions required to get $\lim_{r\to\infty} V(r) = 0$. The minimum is not unique and other sets of parameters can be defined. A more stringent approach is necessary.

\begin{figure*}[ht!]
    \centering
        \includegraphics[scale=0.34]{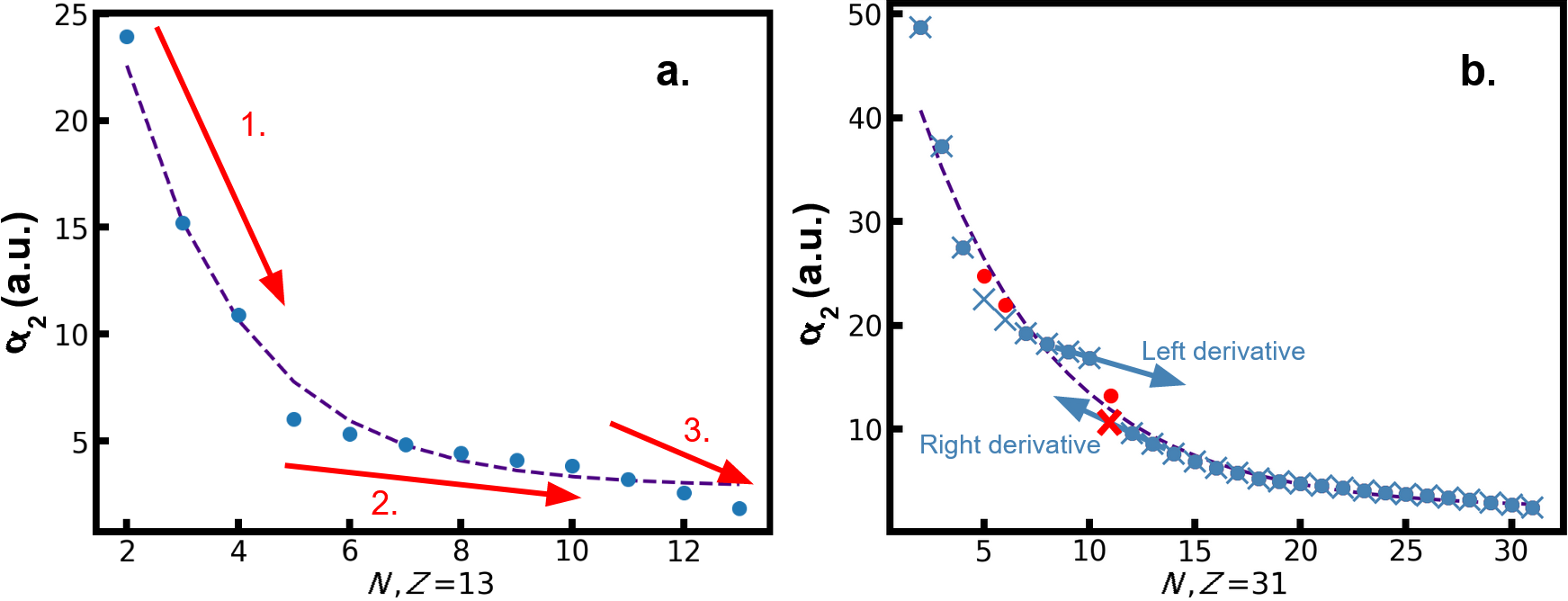}
    \caption{{\bf (a):} Parameter $\alpha_2$ obtained for Al using the exponential fit (dotted line) as initial estimate to minimize the equations over $K_0$, $r_0$ and $r^2_0$. The final values of the parameter do not perfectly coincide the exponential curve. {\bf (b):} Same for Ga. \label{Second_Step}}
\end{figure*}

A second idea consists in imposing three equations for the three parameters of Klapisch 's potential, namely equations on $K_0$, $r_0$ and $r_0^2$. Other moments of $r$ could be used, but we believe that these moments bear more physical meaning, additionally they lead to the excitation energy. As illustrated in Fig. \ref{First_Step} {\bf b.}, We select the parameters of the first minimization satisfying $\alpha_{1,2}>0$, and taking them as an initial estimate, we solve the system of three equations using \verb!scipy.optimize.fsolve!. As before, only the resulting parameters such that $\alpha_{1,2}>0$ are kept. Some of the parameters still need to be defined. It appears that for a fixed $Z_T$, $\alpha_{1,2}$ decreases exponentially with $N$, which allows one to fit the values of the parameters. For the parameter $c$, no explicit shape can be fitted, we only consider the median of the distribution, leading to rough estimates. Consequently, the estimate over the values of $\alpha_{1,2}$ have to be as accurate as possible. Finally, we use the fits to generate an initial estimate for a second minimization over the three equations. For ions with $Z_T<17$, this method gives all three parameters for all $N$. It can be seen, in Fig \ref{Second_Step} {\bf a.}, that the $\alpha_2$ parameter does not perfectly follow the exponential curve. Three distinct areas can be observed: (i) the first 4 electrons added to the ion, following the exponential fit, (ii) the 6 following electrons showing a different behavior, (iii) a new change in the slope when adding more electrons. In fact, it displays changes at the transition between 4 and 5 and then 10 and 11 bound electrons, which correspond to transitions between quantum shells of the atom. This structure echoes the shape of the orbitals: slope variations can be observed at the transitions between orbitals 2s and 2p and then 2p and 3s. As a result, the parameters of Klapisch's potential corresponding to ions with a full valence shell or with only one valence electron sometimes cannot be found with the previously explained method. In that case, in order to estimate their value, we complete either with right or left derivatives or with the mean of the two neighboring values. Depending on the position of the electron, one method can lead to better results than the others, as illustrated in Fig. \ref{Second_Step} {\bf b.} For Ga, the exponential estimate (dotted line) is not sufficient to get acceptable parameters for all $N$ (blue $\bullet$). As a result, a more accurate estimate is required. One can think of completing parameters using the mean value of the closest parameters (red $\bullet$) or the left and right derivative (arrows). In the case of Ga, a guess is required for $N$=11. As the 11$^{th}$ electron is not part of the 2p orbital, only the right derivative bears meaning and leads to a suitable estimate. The final parameters are represented ($\times$). All the parameters of Klapisch's potential are finally properly defined in that way. The values of $K_0$, $r_0$ and $r_0^2$ for each $(Z_T,N)$ perfectly overlap the values given by the GSZ potential. Nevertheless, values of higher-order moments tend to diverge for high values of $N$: the resulting potential is not equivalent to the one of Green, Sellin and Zachor. 

\end{document}